\documentclass[a4paper,11pt]{article}

\usepackage{amssymb,amsmath,amsthm,latexsym,graphicx}
\usepackage{color}
\def\A{\mathcal{A}}
\def\B{\mathcal{B}}
\def\D{\mathcal{D}}
\def\H{\mathcal{H}}
\def\L{\mathcal{L}}
\def\M{\mathcal{M}}
\def\X{\mathcal{X}}
\def\NN{{\mathbb N}}
\newtheorem{definition}{Definition}[section]
\newtheorem{proposition}[definition]{Proposition}
\newtheorem{theorem}[definition]{Theorem}
\newtheorem{lemma}[definition]{Lemma}
\newtheorem{corollary}[definition]{Corollary}
\newtheorem{remark}[definition]{{Remark}}
\newtheorem{example}[definition]{{Example}}

\numberwithin{equation}{section}

\begin{document}
\title{Structure of locally convex quasi $C^*$-algebras}
\author{F. Bagarello, M. Fragoulopoulou, A. Inoue and C. Trapani}
\date{}
\maketitle

\begin{abstract}
The completion of a (normed) $C^*$-algebra $\A_0[\| \cdot \|_0]$ with respect
to a locally convex topology $\tau$ on $\A_0$ that makes the multiplication of
$\A_0$ separately continuous is, in general, a quasi $*$-algebra, and not a
locally convex $*$-algebra \cite{ba-maria-i-t, maria-i-kue}. In this way, one
is led to  consideration of locally convex quasi $C^*$-algebras, which generalize
$C^*$-algebras in the context of quasi $*$-algebras. Examples are given and the
structure of these relatives of $C^*$-algebras is investigated.
\end{abstract}

\section{Introduction}
The study of the structure and representation theory of the
completion of a (normed) $C^*$-algebra $\A_0[\| \cdot \|_0]$ with
respect to a locally convex topology $\tau$ on $\A_0$
"compatible" with the corresponding $^*$-norm topology started in
\cite{ba-maria-i-t} and was continued in \cite{maria-i-kue}. When
the multiplication of $\A_0$ with respect to $\tau$ is jointly
continuous, the completion $\widetilde{\A_0}[\tau]$ of $\A_0[\tau]$
is a $GB^*$-algebra over the unit ball $\mathcal{U}(\A_0) \equiv \{
x \in \A_0: \| x \|_0 \leq 1 \}$ of $\A_0[\| \cdot \|_0]$ if and
only if $\mathcal{U}(\A_0)$ is $\tau$-closed \cite[Corollary
2.2]{maria-i-kue}. When the multiplication of $\A_0$ with respect to
$\tau$ is just separately continuous, $\widetilde{\A_0}[\tau]$ may
fail to be a locally convex $*$-algebra, but may well carry the
structure of a quasi $*$-algebra. The properties and the
$*$-representation theory of $\widetilde{\A_0}[\tau]$, in this case,
have been studied in \cite[Section 3]{ba-maria-i-t} and
\cite[Section 3]{maria-i-kue}. Continuing this project we are led to
the introduction of locally convex quasi $C^*$-algebras in the
present study (see Definition 3.3). In this way, the notion of a
$C^*$-algebra is incorporated within the context of quasi
$*$-algebras. Topological quasi $*$-algebras were first introduced
by G. Lassner (see \cite{lass1,lass2}) for solving problems in
quantum statistics and quantum dynamics that could not be resolved
within the algebraic formulation of quantum theories developed by
Haag and Kastler in \cite{haag-kastler}. However, the bimodule axiom
(which is crucial for many considerations such as $*$-representation
theory) was missing therein and also from many subsequent research papers
for about 20 years! The first correct definition was given in
\cite[p. 90]{schmudgen}, where also large classes of $O^*$-algebra
examples have been derived. Furthermore, quasi
$*$-algebras appeared later in \cite{trapani1,trapani2} and
\cite{ba-t1,ba-t2}. These algebras constitute an interesting class
of the so-called partial $*$-algebras, introduced by J.-P. Antoine
and W. Karwowski in \cite{antoine-karwo1,antoine-karwo2} and studied
extensively in \cite{a-ba-t,ait1,ait2,ba-i-t} and \cite{ait3}.
Partial $*$-algebras and quasi $*$-algebras play an important role
in the theory of unbounded operators, which in its turn has numerous
applications in mathematical physics (see, for instance,
\cite{ait3,inoue,trapani2, bag_rev}).

Our motivation for the present study is clear from the preceding
discussion. The results that we shall exhibit are structured as
follows: After the background material in Section 2, Section 3
defines two notions of positivity in the quasi $*$-algebra
$\widetilde{\A_0}[\tau]$, called "quasi-positivity" and
"commutatively quasi-positivity"; besides, it introduces locally
convex quasi $C^*$-algebras (Definition 3.3) and gives examples from
various classes of topological algebras. Since locally convex quasi
$C^*$-algebras of operators are of particular interest (see, for
instance, Remark 4.2 and Propositions 4.3 and 4.5), we study them
separately in Section 4. In Section 5, the structure of commutative
locally convex quasi $C^*$-algebras is investigated taking into
account \cite[Section 6]{allan1} and
\cite{ba-maria-i-t,maria-i-kue}. In Section 6 we apply the results
of Sections 3 and 5 and also ideas developed in \cite[Section
4]{dixon} and \cite{maria-i-kue} to present a functional calculus
for the quasi-positive elements of a commutative locally convex
quasi $C^*$-algebra. As a consequence the quasi $n$th-root of a
quasi-positive element of such an algebra is, for instance, defined
(Corollary 6.7). In Section 7, the structure of a noncommutative
locally convex quasi $C^*$-algebra is studied. More precisely, if
$\A[\tau]$ is a noncommutative locally convex quasi $C^*$-algebra,
necessary and sufficient conditions are given (see Theorems 7.3 and
7.5) such that $\A[\tau]$ is continuously embedded in a locally
convex quasi $C^*$-algebra of operators. Further, a functional
calculus for commutatively quasi-positive elements in $\A[\tau]$ is
investigated (Theorem 7.8).

\section{Preliminaries}
All algebras that we deal with are complex and the topological
spaces are supposed to be Hausdorff. If an algebra $\A$ has an
identity, this will be denoted by $\mathit{1}$. An algebra $\A$ with
identity $\mathit{1}$, will be called \emph{unital}.

Let $\A_0[\| \cdot \|_0]$ be a $C^*$-algebra. We shall use the
symbol $\| \cdot \|_0$ of the $C^*$-norm to denote the corresponding
topology. Suppose that $\tau$ is a topology on $\A_0$ such that
$\A_0[\tau]$ is a locally convex $*$-algebra. Then, the topologies
$\tau$, $\| \cdot \|_0$ on $\A_0$ are  \textit{compatible} whenever
each Cauchy net in both topologies that converges with respect to
one of them, also converges  with respect to the other one. The
symbol $\widetilde{\A_0}[\tau]$ denotes the completion of
$\A_0[\tau]$.

A \textit{partial $*$-algebra} is a vector space $\A$ equipped with a
vector space involution $*: \A \rightarrow \A: x \mapsto x^*$ and a partial
multiplication defined on a set $\Gamma \subset \A \times \A$ in such a way that:

(i) $(x,y) \in \Gamma$ implies $(y^*, x^*) \in \Gamma$;

(ii) $(x, y_1), (x, y_2) \in \Gamma$ and $\lambda, \mu \in \mathbb{C}$ imply
$(x, \lambda y_1+\mu y_2) \in \Gamma$;

(iii) for every $(x,y) \in \Gamma$, a product $xy \in \A$ is defined, such that
$xy$ depends linearly on $y$ and satisfies  the equality $(xy)^* = y^*x^*$.

Whenever $(x,y) \in \Gamma$, we say that $x$ is a \textit{left
multiplier} of $y$ and $y$ a \textit{right multiplier} of $x$ and we
write $x \in L(y)$, respectively $y \in R(x)$.

Quasi $*$-algebras are important examples of partial $*$-algebras.

If $\A$ is a vector space and $\A_0$ is a subspace of $\A$ such that is also a $*$-algebra,
then $\A$ is said to be a \textit{ quasi $*$-algebra over $\A_0$} whenever the next properties are valid:

(i)$'$ The multiplication of $\A_0$ is extended on $\A$ as follows: The assignments

$\A \times \A_0 \rightarrow \A : (a,x) \mapsto ax \text{ (left multiplication of $x$ by $a$) } $ and

$ \A_0 \times A \rightarrow \A: (x,a) \mapsto xa \text{ (right multiplication of $x$ by $a$) }$\\
are always defined and are bilinear;

(ii)$'$ $x_1(x_2 a) = (x_1 x_2) a, (a x_1)x_2= a(x_1 x_2)$ and $x_1 (a x_2) = (x_1 a ) x_2$,
for all $x_1, x_2 \in \A_0$ and $a \in \A$;

(iii)$'$ the involution $*$ of $\A_0$ is extended on $\A$, denoted
also by $*$, such that $(ax)^* = x^* a^*$ and $(xa)^* = a^* x^*$,
for all $x \in \A_0$ and $a \in \A$.

For further information see \cite{ait3}. If $\A_0[\tau]$ is a
locally convex $*$-algebra, with separately continuous
multiplication, its completion $\widetilde{\A_0}[\tau]$ is a quasi
$*$-algebra over $\A_0$ under the following operations: Given $x \in
\A_0$ and $a \in \widetilde{\A_0}[\tau]$

$\bullet$ \ \ $ax := \displaystyle \lim_\alpha x_\alpha x \text{
(left multiplication) }$

$\bullet$ \ \ $xa := \displaystyle \lim_\alpha x x_\alpha \text{ (right multiplication) }$ \\
with $\{x_\alpha\}_{\alpha \in \Delta}$ a net in $\A_0$ such that
$a = \tau$-$\displaystyle \lim_\alpha x_\alpha$.

$\bullet$ \ \ An involution on $\widetilde{\A_0}[\tau]$ like in (iii)$'$ is
the continuous extension of the involution given on $\A_0$.

A $*$-invariant subspace $\A$ of $\widetilde{\A_0}[\tau]$ containing
$\A_0$ is called a \textit{quasi $*$-subalgebra} of
$\widetilde{\A_0}[\tau]$ if $ax$, $xa$ belong to $\A$ for any $x \in
\A_0$, $a \in \A$. Then, one easily shows that $\A$ is a quasi
$*$-algebra over $\A_0$. Moreover, $\A[\tau]$ is a locally convex
space that contains $\A_0$ as a dense subspace and for every fixed
$x \in \A_0$, the maps $\A[\tau] \rightarrow \A[\tau]$ with $a
\mapsto ax$ and $a \mapsto xa$ are continuous. An algebra of this
kind is called \textit{locally convex quasi $*$-algebra over
$\A_0$}.

Another concept we need is that of a $GB^*$-algebra introduced by
G.R. Allan in 1967 \cite{allan2} for generalizing $C^*$-algebras
(also see \cite{dixon}). Let $\A[\tau]$ be a unital locally convex
$*$-algebra. Let $\B^*$ be the collection of all closed, bounded,
absolutely convex subsets $B$ of $\A[\tau]$ with the properties:
$\mathit{1} \in B, B^* = B$ and $B^2 \subset B$. For every $B \in
\B^*$, the linear span $A[B]$ of $B$ is a normed $*$-algebra under
the Minkowski functional $\| \cdot \|_B$ of $B$. If $A[B]$ is
complete for every $B \in \B^*$, then $\A[\tau]$ is said to be
\textit{ pseudo-complete}. Every sequentially complete locally
convex $*$-algebra is pseudo-complete \cite[Proposition
(2.6)]{allan1}. Now, a unital pseudo-complete locally convex
$*$-algebra $\A[\tau]$, such that $\B^*$ has a greatest member,
denoted by $B_0$, and $(\mathit{1}+x^*x)^{-1}$ exists and belongs to
$A[B_0]$ for every $x \in \A$, is called a \textit{$GB^*$-algebra
over $B_0$}. In this case $A[B_0]$ is a $C^*$-algebra.

\section{ Locally convex quasi $C^*$-algebras}
Throughout this Section $\A_0[\| \cdot \|_0]$ denotes a unital
$C^*$-algebra and $\tau$ a locally convex topology on $\A_0$
compatible with the corresponding $\| \cdot \|_0$-topology. Under
certain conditions on $\tau$ a quasi $*$-subalgebra $\A$ of the
quasi $*$-algebra $\widetilde{\A_0}[\tau]$ over $\A_0$ is formed,
which is named locally convex quasi $C^*$-algebra. Examples and
basic properties of such algebras are presented. So, let $\A_0[\|
\cdot \|_0]$ and $\tau$ be as above with $\{p_\lambda\}_{\lambda \in
\Lambda}$ a defining family of seminorms for $\tau$. Suppose that
$\tau$ satisfies the properties:

(T$_1$) $\A_0[\tau]$ is a locally convex $*$-algebra with separately continuous multiplication.

(T$_2$) $\tau \preceq \| \cdot \|_0$.\\
Then, the identity map $\A_0[\| \cdot \|_0] \rightarrow \A_0[\tau]$
extends to a continuous $*$-linear map $\A_0[\| \cdot \|_0]
\rightarrow \widetilde{\A_0}[\tau]$ and since $\tau, \| \cdot \|_0$
are compatible, the $C^*$-algebra $\A_0[\| \cdot \|_0]$ can be
regarded embedded into $\widetilde{\A_0}[\tau]$. It is easily shown
that $\widetilde{\A_0}[\tau]$ is a quasi $*$-algebra over $\A_0$
(cf. \cite[Section 3]{maria-i-kue}).

The next Definition 3.1 provides concepts of positivity for elements of a
quasi $*$-algebra $\widetilde{\A_0}[\tau]$.

\begin{definition} {\rm
An element $a$ of $\widetilde{\A_0}[\tau]$ is called
\textit{quasi-positive} (resp. \textit{commutatively
quasi-positive}) if there is a net (resp. commuting net)
$(x_\alpha)_{\alpha \in \Delta}$ of the positive cone $(\A_0)_+$ of
the $C^*$-algebra $\A_0[\| \cdot \|_0]$, which converges to $a$ with
respect to the topology $\tau$. }
\end{definition}

We have already used the symbol $(\A_0)_+$ for the set of all
positive elements of the $C^*$-algebra $\A_0[\| \cdot \|_0]$. The
set of all quasi-positive (resp. commutatively quasi-positive)
elements of $\widetilde{\A_0}[\tau]$, we shall denote by
$\widetilde{\A_0}[\tau]_{q+}$ (resp.
$\widetilde{\A_0}[\tau]_{cq+}$). Then, $\widetilde{\A_0}[\tau]_{q+}$
is a wedge (that is, for any $a,b \in \widetilde{\A_0}[\tau]_{q+}$
and $\lambda \geq 0$, the elements $a+b$ and $\lambda a$ belong to
$\widetilde{\A_0}[\tau]_{q+}$), but it is not necessarily a positive
cone (i.e. $\widetilde{\A_0}[\tau]_{q+} \cap
(-\widetilde{\A_0}[\tau]_{q+}) \neq \{ 0 \}$). The set
$\widetilde{\A_0}[\tau]_{cq+}$ is not even, in general, a wedge.
But, if $\A_0$ is commutative, then of course,
$\widetilde{\A_0}[\tau]_{q+} = \widetilde{\A_0}[\tau]_{cq+}$.

Further, we employ the following two extra conditions (T$_3$), (T$_4$) for the locally
convex topology $\tau$ on $\A_0$ and examine the effect on $\widetilde{\A_0}[\tau]_{cq+}$:

(T$_3$) For each $\lambda \in \Lambda$, there exists $\lambda' \in \Lambda$ such that
\[
p_\lambda(xy) \leq \| x \|_0 p_{\lambda'}(y), \forall \ x, y \in
\A_0 \text{ with } xy = yx;
\]

(T$_4$) The set $\mathcal{U}(\A_0)_+ := \{ x \in (\A_0)_+ : \| x
\|_0 \leq 1 \}$ is $\tau$-closed, and

\hspace{0,7cm} $\widetilde{\A_0}[\|\cdot\|]_{q+} \cap \A_0 =
(\A_0)_+$.

\

\begin{proposition}
Let $\A_0[\| \cdot \|_0]$ be a unital $C^*$-algebra and $\tau$ a
locally convex topology on $\A_0$. Suppose that $\tau$ fulfils the
conditions \emph{(T$_1$)-(T$_4$)}. Then, $\widetilde{\A_0}[\tau]$ is
a locally convex quasi $*$-algebra over $\A_0$ with the properties:

\emph{(1)} For every $a \in \widetilde{\A_0}[\tau]_{cq+}$, the element $\mathit{1}+a$
is invertible and its inverse $(\mathit{1}+a)^{-1}$ belongs to $\mathcal{U}(\A_0)_+$.

\emph{(2)} For a given $a \in \widetilde{\A_0}[\tau]_{cq+}$ and any $\varepsilon >0$, let
\[
a_\varepsilon= a (\mathit{1}+\varepsilon a)^{-1}.
\]
Then, $\{a_\varepsilon\}_{\varepsilon>0}$ is a commuting net in $(\A_0)_+$ such that
$a- a_\varepsilon \in \widetilde{\A_0}[\tau]_{cq+}$ and $a=
\tau$-$\displaystyle \lim_{\varepsilon \rightarrow 0} a_\varepsilon$.

\emph{(3)} $\widetilde{\A_0}[\tau]_{cq+} \cap (-\widetilde{\A_0}[\tau]_{cq+}) = \{ 0 \}$.

\emph{(4)} If $a \in \widetilde{\A_0}[\tau]_{cq+}$ and $b \in (\A_0)_+$ such that
$b-a \in \widetilde{\A_0}[\tau]_{q+}$, then $a \in (\A_0)_+$.
\end{proposition}
\begin{proof}
(1) Let $a \in \widetilde{\A_0}[\tau]_{cq+}$. Then, there is a net
$\{x_\alpha \}_{\alpha \in \Delta}$ in $(\A_0)_+$, such that
$x_\alpha x_\beta = x_\beta x_\alpha$, for all $\alpha, \beta \in
\Delta$, and $x_\alpha \xrightarrow[\tau]{}a$. Using properties of
the positive elements of a $C^*$-algebra and the condition (T$_3$),
we have that for every $\lambda \in \Lambda$, there is $\lambda' \in
\Lambda$ with
\begin{align*}
p_\lambda((\mathit{1}+x_\alpha)^{-1}- ( \mathit{1} + x_\beta)^{-1})
&= p_\lambda ((\mathit{1}+x_\alpha)^{-1}(x_\alpha-x_\beta)(\mathit{1}+x_\beta)^{-1}) \\
&\leq \| (\mathit{1}+x_\alpha)^{-1}\|_0 \| (\mathit{1}+x_\beta)^{-1}\|_0 p_{\lambda'}(x_\alpha - x_\beta) \\
& \leq p_{\lambda'}(x_\alpha - x_\beta) \rightarrow 0.
\end{align*}
So, $\{(\mathit{1}+x_\alpha)^{-1}\}_{\alpha \in \Delta}$ is a Cauchy net in $\A_0[\tau]$
consisting of elements of $\mathcal{U}(\A_0)_+$, which by (T$_4$) is $\tau$-closed.
Hence,
\begin{align}
(\mathit{1}+x_\alpha)^{-1} \xrightarrow[\tau]{}y \in \mathcal{U}(\A_0)_+.
\end{align}
We shall show that $(\mathit{1}+a)^{-1}$ exists and equals $y$.
Indeed: Using again condition (T$_3$), for each $\lambda \in \Lambda$, there is $\lambda' \in \Lambda$ with
\begin{align*}
p_\lambda ( \mathit{1}-(\mathit{1}+a)(\mathit{1}+x_\alpha)^{-1})
&= p_\lambda ((x_\alpha - a)(\mathit{1}+x_\alpha)^{-1}) \\
&\leq \| (\mathit{1}+x_a)^{-1} \|_0 p_{\lambda'} (x_\alpha-a) \leq p_{\lambda'}(x_\alpha - a) \rightarrow 0.
\end{align*}
Therefore,
\begin{align}
(\mathit{1}+a) (\mathit{1}+x_\alpha)^{-1}
\xrightarrow[\tau]{} \mathit{1}.
\end{align}
On the other hand, since
\[
x_\beta y = \tau-\lim_\alpha x_\beta (\mathit{1} +x_\alpha)^{-1}
          = \tau-\lim_\alpha (\mathit{1} +x_\alpha)^{-1} x_\beta
          = y x_\beta, \ \forall \ \beta \in \Delta,
\]
 we have $ay=ya$. Further, we can show that
\begin{align}
  (\mathit{1} +a)(\mathit{1} +x_\alpha)^{-1} \xrightarrow[\tau]{} (\mathit{1} +a)y.
\end{align}
Indeed, since $x_\alpha \xrightarrow[\tau]{}a$, for any $\varepsilon
> 0$ there exists $\alpha_0 \in \Delta$ such that for all $\alpha \geq \alpha_0$ and
all $\lambda \in \Lambda$ one has $p_{\lambda} (x_\alpha - a) <
\varepsilon$. Now, by (T$_3$) we have that for any $\alpha \in
\Delta$
\begin{align*}
&p_\lambda((\mathit{1} + a)(\mathit{1} +x_\alpha)^{-1} - (\mathit{1} + a)y)\\
& \leq p_\lambda ((\mathit{1} + a)(\mathit{1} +x_\alpha)^{-1} - (\mathit{1}
+x_{\alpha_0})(\mathit{1} +x_\alpha)^{-1}) \\
& + p_\lambda ((\mathit{1} + x_{\alpha_0})(\mathit{1} +
x_\alpha)^{-1} - (\mathit{1} + x_{\alpha_0})y) +
p_\lambda ((\mathit{1} + x_{\alpha_0})y -(1 + a)y) \\
& \leq p_{\lambda'}(a - x_{\alpha_0}) +\|\mathit{1} + x_{\alpha_0}\|_0 p_\lambda'((\mathit{1}
 +x_\alpha)^{-1} - y) +p_\lambda (x_{\alpha_0} - a) \\
& < 2\varepsilon + \|\mathit{1} + x_{\alpha_0}\|_0 p_\lambda'((\mathit{1} +x_\alpha)^{-1} - y),
\end{align*}
which by (3.1) implies that $\lim_\alpha p_\lambda((\mathit{1} + a)(\mathit{1}
+x_\alpha)^{-1} - (\mathit{1} + a)y) = 0$.
 Thus,  by (3.2) and (3.3) we have $(\mathit{1} + a)y= \mathit{1} =y(\mathit{1} +
 a)$.
Hence, $(\mathit{1}+a)^{-1}$ exists and belongs to
$\mathcal{U}(\A_0)_+$ (since $y$ does).

(2) It is clear from (1) that for every $\varepsilon >0$ the element
$(\mathit{1}+\varepsilon a)^{-1}$ exists in $\widetilde{\A_0}[\tau]$ and belongs
to $\mathcal{U}(\A_0)_+$. In particular, applying (T$_3$) we get that for each
$\lambda \in \Lambda$, there is $\lambda' \in \Lambda$ with
\[
p_\lambda(\mathit{1}-(\mathit{1}+\varepsilon a)^{-1}) = \varepsilon
p_\lambda (a(\mathit{1}+\varepsilon a)^{-1}) \leq \varepsilon \|
(\mathit{1}+\varepsilon a)^{-1}\|_0 p_{\lambda'}(a) \leq \varepsilon
p_{\lambda'}(a),
\]
so that
\begin{equation}
\tau \text{-}\displaystyle \lim_{\varepsilon \rightarrow 0} (\mathit{1}+ \varepsilon a)^{-1} = \mathit{1}.
\end{equation}

On the other hand, from the very definitions one has
\[
a_\varepsilon = a(\mathit{1}+\varepsilon a)^{-1} = (\mathit{1}+
\varepsilon a)^{-1}a= \dfrac{1}{\varepsilon}
(\mathit{1}-(\mathit{1}+\varepsilon a)^{-1}), \quad \forall \
\varepsilon > 0, \ \text{ and}
\]
\begin{equation}
a-a_\varepsilon = a(\mathit{1}-(\mathit{1}+ \varepsilon a)^{-1})=
(\mathit{1}-(\mathit{1}+ \varepsilon a)^{-1}) a \in \widetilde{\A_0}[\tau]_{cq+}.
\end{equation}
Now, by the same way as in (3.3), we conclude from (3.4) and (3.5)
that $\tau \text{-}\displaystyle \lim_{\varepsilon \rightarrow
0}a_\varepsilon = a$.

(3) Let $a \in \widetilde{\A_0}[\tau]_{cq+} \cap (-\widetilde{\A_0}[\tau]_{cq+})$.
For any $\varepsilon>0$, we have by (2) that
\[
(\A_0)_+ \ni a(\mathit{1}+ \varepsilon a)^{-1} \xrightarrow[\tau]{} a
\text{ and } (\A_0)_+ \ni (-a)(\mathit{1}-\varepsilon a)^{-1} \xrightarrow[\tau]{} -a.
\]
Thus, if
\begin{equation}
x_\varepsilon := a( \mathit{1}+ \varepsilon a)^{-1} - (-a) (\mathit{1}- \varepsilon a)^{-1},
\end{equation}
we get
\begin{align*}
x_\varepsilon = a((\mathit{1}+ \varepsilon a)^{-1} + (\mathit{1}- \varepsilon a)^{-1}) &=
a( \mathit{1} + \varepsilon a)^{-1}(\mathit{1}- \varepsilon a+ \mathit{1}+
\varepsilon a)(\mathit{1}- \varepsilon a)^{-1} \\
&= 2a ( \mathit{1}+ \varepsilon a)^{-1}(\mathit{1}- \varepsilon a)^{-1},
\end{align*}
where by (1) and (2) we conclude that $(\mathit{1}-\varepsilon
a)^{-1} \in (\A_0)_+$ and $a ( \mathit{1} + \varepsilon a)^{-1} \in
(\A_0)_+$ respectively. Therefore, $x_\varepsilon  \in (\A_0)_+$
according to the functional calculus in commutative $C^*$-algebras.
Similarly, we have that
\[
-x_\varepsilon = 2(-a) (\mathit{1}-\varepsilon a)^{-1} ( \mathit{1}+ \varepsilon a)^{-1} \in (\A_0)_+
\]
since $(-a)(\mathit{1}-\varepsilon a)^{-1}$ and $(\mathit{1}+\varepsilon a)^{-1}$ belong to $(\A_0)_+$.
Thus,
\[
x_\varepsilon \in (\A_0)_+ \cap (-(\A_0)_+)= \{ 0 \}
\]
and so (see (3.6))
\[
a(\mathit{1}+\varepsilon a)^{-1} = -a(\mathit{1}-\varepsilon a)^{-1}.
\]
Taking $\tau$-limits with $\varepsilon \rightarrow 0$, we get $a= -a$, i.e., $a=0$.

(4) By (2) and the assumptions in (4), $b-a$ and $a-a_\varepsilon$
are contained in $\widetilde{\A_0}[\tau]_{q+}$. Since,
$\widetilde{\A_0}[\tau]_{q+}$ is a wedge, $b-a_\varepsilon =
(b-a)+(a-a_\varepsilon) \in \widetilde{\A_0}[\tau]_{q+}$.
Furthermore, by (T$_4$)
\[
b-a_\varepsilon \in \widetilde{\A_0}[\tau]_{q+} \cap \A_0 =
(\A_0)_+, \quad \forall \ \varepsilon > 0.
\]
Hence,
\[
\| a_\varepsilon \|_0 \leq \| b \|_0, \quad \forall \ \varepsilon >
0,
\]
so that if $b=0$, then $a= 0 \in (\A_0)_+$ since
$a=\tau$-$\displaystyle \lim_{\varepsilon \rightarrow 0}
a_\varepsilon$. If $b\neq 0$ then $\left\{ \frac{a_\varepsilon}{\|
b\|_0} : \varepsilon>0 \right\} \subset \mathcal{U}(\A_0)_+$ and by
(T$_4$) $\mathcal{U}(\A_0)_+$ is $\tau$-closed; so again we get that
$a \in (\A_0)_+$.
\end{proof}

The above lead to the following

\begin{definition}{\rm
A quasi $*$-subalgebra $\A$ of the locally convex quasi $*$-algebra
$\widetilde{\A_0}[\tau]$ over $\A_0$, where $\A_0[\| \cdot \|_0]$ is
a unital $C^*$-algebra and $\tau$ a locally convex topology on
$\A_0$ satisfying the conditions (T$_1$)-(T$_4$), is said to be a
\textit{locally convex quasi $C^*$-algebra over $\A_0$}. }
\end{definition}

We present now some examples of locally convex quasi $C^*$-algebras.

\begin{example}[$GB^*$-algebras] {\rm
Let $\A[\tau]$ be a $GB^*$-algebra over $B_0$ (see Section 2).
Then, $\A_0[\| \cdot \|_0] \equiv A[B_0]$ is a $C^*$-algebra under the
$C^*$-norm $\| \cdot \|_0 \equiv \| \cdot \|_{B_0}$ given by the Minkowski functional of $B_0$.
Assume that the locally convex topology $\tau$ fulfils the condition (T$_3$).
Then, it is easily checked that $\A[\tau]$ is a locally convex quasi $C^*$-algebra over $\A_0$. }
\end{example}

\begin{example}[Banach quasi $C^*$-algebras] {\rm
Let $\A_0[\| \cdot \|_0]$ be a unital $C^*$-algebra and $\tau= \|
\cdot \|$ a norm topology on $\A_0$ with the properties
(T$_1$)-(T$_4$). That is,

(T$_1$) $\A_0[ \| \cdot \|] $ is a locally convex $*$-algebra;

(T$_2$) $\| \cdot \| \preceq \| \cdot \|_0$;

(T$_3$) $\| xy \| \leq \|x\|_0 \| y \|, \quad \forall \ x,y \in
\A_0$ with $xy = yx$;

(T$_4$) $\mathcal{U}(\A_0)_+$ is $\| \cdot \|$-closed, and
$\widetilde{\A_0}[\|\cdot\|]_{q+} \cap \A_0 = (\A_0)_+$. \\
Then, a locally convex quasi $C^*$-algebra over $\A_0$ is called a
\textit{normed quasi $C^*$-algebra over} $\A_0$.
In particular, the completion $\widetilde{\A_0}[\| \cdot \|]$ of $\A_0[\| \cdot \|]$
is said to be a \textit{Banach quasi $C^*$-algebra over} $\A_0$.

Notice that the Banach space $L^p[0,1]$, $1 \leq p < \infty$, is a Banach quasi
$C^*$-algebra over the $C^*$-algebra $L^\infty[0,1]$. }
\end{example}

\begin{example}[proper $CQ^*$-algebras]{\rm
A quasi $*$-algebra $(\X, \A_0)$ is said to be a \textit{Banach quasi $*$-algebra}
over $\A_0$ (see \cite{ba-t1}), if a norm $\| \cdot \|$ is defined on $\X$ with the properties:

(i) $\X[ \| \cdot \| ]$ is a Banach space;

(ii) $\| x^* \| = \| x \|, \ \forall \ x \in \X$;

(iii) $\A_0$ is dense in $\X[ \| \cdot \|]$;

(iv) for each $a \in \A_0$, the map $L_a: \X \rightarrow \X: x \mapsto ax$, is continuous.

The continuity of the involution implies that for each $a \in \A_0$, the map $R_a :
\X \rightarrow \X: x \mapsto xa$, is continuous.

The identity of $(\X, \A_0)$ is an element $\mathit{1} \in \A_0$
such that $\mathit{1} x = x  \mathit{1} = x$, for each $x \in \X$.
Let $(\X, \A_0)$ be a unital Banach quasi $*$-algebra. Then, $\A_0$
is a normed $*$-algebra under the norm
\[
\|a\|_{op}:= \max{ \{ \| L_a\|, \|R_a \| \}}, \quad \forall \ a \in
\A_0, \text{ and }
\]
\begin{align}
&\| a \| \leq \| a \|_{op}, \quad \forall \ a \in \A_0 , \\
& \|ab \| \leq \| a \| \| b \|_{op}, \quad \forall \ a,b \in \A_0.
\end{align}
An element $x$ of $\X$ is said to be \textit{bounded} if the map
$R_x: \A_0 \rightarrow \X: a \mapsto ax$ is continuous, equivalently
the map $L_x: \A_0 \rightarrow \X: a \mapsto xa$ is continuous.
Then, $R_x$ respectively $L_x$ extend to bounded linear operators
$\overline{R}_x$ resp. $\overline{L}_x$. Denote by $\X_b$ the set of
all bounded elements of $\X$. Then $\X$ is said to be
\textit{normal} \cite{trapani3} if $\overline{L}_x y= \overline{R}_y
x$ for every $x,y \in \X_b$. In this case, $\X_b$ is a Banach
$*$-algebra equipped with the multiplication
\[
x \circ y = \overline{L}_x y,\quad \forall \ x,y \in \X_b
\]
and the norm $\| x \|_b:= \max{\{ \| \overline{L}_x \|, \| \overline{R}_x \| \}}, x \in \X_b$ (see
\cite[Corollary 2.14]{trapani3}). Furthermore, we have
\begin{equation}
{\overline{\mathcal{U}(\A_0[\| \cdot \|_{op}])}}^{\| \cdot \|} \subset \mathcal{U}(\X_b).
\end{equation}
Indeed, take an arbitrary $x \in {\overline{\mathcal{U}(\A_0[\|
\cdot \|_{op}])}}^{\| \cdot \|} $. Then, there is a sequence $\{ a_n
\}$ in $\mathcal{U}(\A_0[ \| \cdot \|_{op}])$ such that
$\displaystyle \lim_{n \rightarrow \infty} \| a_n -x \|=0$. On the
other hand, using (3.8), we have that for each $b \in \A_0$
\[
\|xb \| = \lim_n \| a_n b \| \leq \varlimsup_{n \rightarrow \infty} \| a_n \|_{op} \| b \| \leq \|b\|
\]
and similarly $\| bx \| \leq \|b \|$. Hence, $x \in
\mathcal{U}(\X_b)$.}
\end{example}

If $\A_0 = \X_b$, then the Banach quasi $*$-algebra $(\X, \A_0)$ is
said to be \textit{full}. If $\A_0[ \| \cdot \|_{op}]$ is a
$C^*$-algebra, then $(\X, \A_0)$ is called a \textit{proper
$CQ^*$-algebra} \cite{ba-t1}.

Let $(\X, \A_0 )$ be a full proper $CQ^*$-algebra. Suppose
$\widetilde{\A_0} [\|\cdot\|]_{q+} \cap \A_0  = (\A_0 )_+$. Then,
$\mathcal{U}(\A_0)_+$ is $\| \cdot \|$-closed. Indeed, take an
arbitrary $x \in \overline{\mathcal{U}(\A_0)_{+}} ^{\| \cdot \|} $.
Then, there is a sequence $\{ a_n \}$ in $\mathcal{U}(\A_0)_+$ such
that $\displaystyle \lim_{n \rightarrow \infty} \| a_n -x \|=0$.
Since $(\X, \A_0 )$ is full, it follows from (3.9) that $x \in
\mathcal{U}(\A_0) $, which implies $x \in
\widetilde{\A_0}[\|\cdot\|]_{q+} \cap \A_0 = (\A_0)_+$. Thus,
$\mathcal{U}(\A_0)_{+}$ is $\|\cdot\|$-closed.

Banach quasi $C^*$-algebras are related to proper $CQ^*$-algebras in the following way:

{\bf 1}. \emph{ If $(\X, \A_0)$ is a full proper $CQ^*$-algebra with
$\widetilde{\A_0}[\|\cdot\|]_{q+} \cap \A_0 = (\A_0)_+$, then $\X$
is a Banach quasi $C ^*$-algebra over the $C^*$-algebra $\A_0[ \|
\cdot \|_{op}]$.}
\\
\indent This follows by the very definitions (in this respect, see
also Example 3.5) and (3.7), (3.8), (3.9).

{\bf 2}.
Conversely, \emph{suppose that $\A$ is a Banach quasi $C^*$-algebra over the $C^*$-algebra
$\A_0[ \| \cdot \|_0]$. Then, $(\A, \A_0)$ is a proper $CQ^*$-algebra if and only if
$\|a\|_{op} = \|a \|_0$, for all $a \in \A_0$.}

\medskip

We consider the following realization of this situation. Let $I$ be
a compact interval of $\mathbb{R}$. Then, it is shown that the
proper $CQ^*$-algebra $(L^p(I), L^\infty(I))$ is a Banach quasi
$C^*$-algebra over $L^\infty(I)$, but the proper $CQ^*$-algebra
$(L^p(I), C(I))$ is not a Banach quasi $C^*$-algebra over $C(I)$.
\medskip

A noncommutative example of a proper $CQ^*$-algebra, which is also a
Banach quasi $C^*$-algebra, can be constructed as follows. Let $S$
be a (possibly unbounded) selfadjoint operator in a Hilbert space
$\H$, with $S\geq I$. Let ${\mathcal C}(S)$ be the von Neumann
algebra
$${\mathcal
C}(S)=\{X\in \B(\H):\, XS^{-1}=S^{-1}X\},$$ where $\B(\H)$ is the
$C^*$-algebra of all bounded linear operators on $\H$. We denote
with $\| \cdot \|_0$ the operator norm in $\B(\H)$. Let us define on
${\mathcal C}(S)$ the norm
$$ \|X\|= \|S^{-1}XS^{-1}\|_0, \quad X \in {\mathcal C}(S) .$$
 Let $\widetilde{{\mathcal C}(S)}$ denote
the $\|\cdot \|$-completion of ${\mathcal C}(S)$. Then, it is easily
seen that $(\widetilde{{\mathcal C}(S)},{\mathcal C}(S))$ is a
proper $CQ^*$-algebra. Making use of the weak topology of $\B(\H)$,
one can prove that (T$_4$) also holds on ${\mathcal C}(S)$. The
proof will be given in the next Section in a more general context.
Then, $\widetilde{{\mathcal C}(S)}$ is a locally convex quasi
$C^*$-algebra.
\medskip

\begin{example}{\rm In this example we will shortly
discuss the so-called physical topologies on a noncommutative
$C^*$-algebra, first introduced by Lassner \cite{lass1, lass2} in
the early 1980's. Thereafter these topologies revealed to be very
useful for the description of many quantum physical models with an
infinite number of degrees of freedom (for reviews see
\cite{trapani2, bag_rev} and\cite[Ch. 11]{ait3}). In view of these
applications, it seems interesting to consider the question under
which conditions they can be cast in the framework developed in this
paper.

Let $\A_0$ be a $C^*$-algebra and $\Sigma=\{\pi_\alpha; \alpha \in
I\}$ a system of $*$-representations of $\A_0$ on a dense subspace
$\D_\alpha$ of a Hilbert space $\H_\alpha$, i.e. each $\pi_\alpha$
is a $*$-homomorphism of $\A_0$ into the $O^*$-algebra
$\L^\dag(\D_\alpha)$ (see Section 4). Since $\A_0$ is a
$C^*$-algebra, each $\pi_\alpha$ is a bounded $*$-representation,
i.e. $\overline{\pi_\alpha(x)}\in \B(\H_\alpha)$, for every $x \in
\A_0$. The system $\Sigma$ is supposed to be {\em faithful}, in the
sense that if $x \in \A_0$, $x \neq 0$, then there exists $\alpha
\in \Sigma$ such that $\pi_\alpha(x)\neq 0$. The physical topology
$\tau_\Sigma$ is the coarsest locally convex topology on $\A_0$ such
that every $\pi_\alpha \in \Sigma$ is continuous from
$\A[\tau_\Sigma]$ into
$\L^\dag(\D_\alpha)[\tau_u(\L^\dag(\D_\alpha))]$, where
$\tau_u(\L^\dag(\D_\alpha))$ is the $\L^\dag(\D_\alpha)$-uniform
topology of $\L^\dag(\D_\alpha)$ (see Section 4). This topology
depends, of course, on the choice of an appropriate system $\Sigma$
of $*$-representations of $\A_0$; these  $*$-representations are, in
general nothing but the $GNS$ representations constructed starting
from a family $\omega_\alpha$ of states which are {\em relevant}
(and they are usually called in this way) for the physical model
under consideration. Every physical topology satisfies the
conditions (T$_1$), (T$_2$) and (T$_4$), but it does not necessarily
satisfy (T$_3$). Here we show that $\widetilde{\A_0}[\tau_\Sigma]$
is a locally convex quasi $C^*$-algebra over $\A_0$ for some special
choice of the system $\Sigma$ of *-representations of $\A_0$.
Suppose that $\D_\alpha= \D^\infty(M_\alpha)=\bigcap_{n \in
\NN}\D(M_\alpha^n)$, where $M_\alpha$ is a selfadjoint unbounded
operator. Without loss of generality we may assume that
$M_\alpha\geq I_\alpha$, with $I_\alpha$ the identity operator in
$\B(\H_\alpha)$. Let $\Sigma$ be a system of representations
$\pi_\alpha$ of $\A_0$ on $\D_\alpha$ such that
$\pi_\alpha(x)M_\alpha\xi= M_\alpha\pi_\alpha(x)\xi$, for every $x
\in \A_0$ and for every $\xi \in \D_\alpha$. Then
$\widetilde{\A_0}[\tau_\Sigma]$ is a locally convex quasi
$C^*$-algebra over $\A_0$.  This follows from the fact that, in this
case, the physical topology $\tau_\Sigma$ is defined by the family
of seminorms
$$ p_\alpha^f (x) := \|f(M_\alpha)\overline{\pi_\alpha(x)}\|_0 \
\text{(operator $C^*$-norm)}, \ \forall \ x \in \A_0, $$ where
$\pi_\alpha\in \Sigma$  and $f$ runs over the set ${\mathcal F}$ of
all positive, bounded and continuous functions on ${\mathbb R}^+$
such that $ \sup_{x \in {\mathbb R}^+} x^kf(x)<\infty$, for every $
\ k=0,1,2,\ldots$ \cite[Lemma 2.8]{lass2}, and from the inequality
$$ p_\alpha^f (xy)=
\|f(M_\alpha)\overline{\pi_\alpha(x)}\overline{\pi_\alpha(y)}\|_0\leq
\|\overline{\pi_\alpha(x)}\|_0p_\alpha^f (y), \ \forall \ x,y \in
\A_0.$$}

\end{example}

\section{Locally convex quasi $C^*$-algebras of operators}
Let $\D$ be a dense subspace in a Hilbert space $\H$.
Let $\L(\D)$ be the algebra (under usual algebraic operations) of all linear operators
from $\D$ to $\D$ and $\L^\dag(\D):= \{ X \in \L(\D): \D(X^*) \supset \D \text{ and }
X^* \D \subset \D \}$, where $\D(X^*)$ stands for the domain of the adjoint $X^*$ of $X$.
Then $\L^\dag(\D)$ is a $*$-algebra under the involution $X^\dag := X^* \lceil \D$
(see \cite[p.8]{inoue}).
Furthermore, let $\L^\dag(\D, \H)$ denote all linear operators $X$ from $\D$ to $\H$
such that $\D(X^*) \supset \D$.
Then, $\L^\dag(\D,\H)$ is a $*$-preserving vector space endowed with the usual
linear operations and the involution $X^\dag:= X^* \lceil \D$ (ibid., p.23).
In particular, $\L^\dag(\D, \H)$ is a partial $*$-algebra \cite[Proposition 2.1.11]{ait3}
under the (weak) partial multiplication $X \Box Y = X^{\dag *}Y$, defined whenever
$Y \D \subset \D(X^{\dag *})$ and $X^\dag \D \subset \D(Y^*), X, Y \in \L^\dag(\D,\H)$.

Let now $\M_0$ be a unital $C^*$-algebra over $\H$ that leaves $\D$
invariant, i.e., $\M_0 \D \subset \D$. Then, the restriction
$\M_0\lceil \D$ of $\M_0$ to $\D$ is an $O^*$-algebra on $\D$,
therefore an element $X$ of $\M_0$ is regarded as an element $X
\lceil \D$ of $\M_0 \lceil \D$. Moreover, let
\[
\M_0 \subset \M \subset \L^\dag(\D, \H),
\]
where $\M$ is an $O^*$-vector space on $\D$, that is, a
$*$-invariant subspace of $\L^\dag(\D, \H)$. Denote by $\B(\M)$ the
set of all bounded subsets of $\D[t_\M]$, where $t_\M$ is the graph
topology on $\M$ (see \cite[p.9]{inoue}). Further, denote by
$\B_f(\D)$ the set of all finite subsets of $\D $. Then $\B_f(\D)
\subset \B(\M)$. A subset $\B$ of $\B(\M)$ is called
\emph{admissible} if the following hold:

(i) $\B_f(\D) \subset \B$,

(ii) $\forall \ \mathfrak{M}_1, \mathfrak{M}_2 \in \B, \ \exists \
\mathfrak{M}_3 \in \B : \mathfrak{M}_1 \cup \mathfrak{M}_2 \subset
\mathfrak{M}_3$,

(iii) $A\mathfrak{M} \in \B, \ \forall \ A \in \M_0 \
\text{and } \  \forall \ \mathfrak{M} \in \B.$ \\
It is clear that $\B_f(\D)$ and $\B(\M)$ are admissible. Consider
now an arbitrary admissible subset $\B$ of $\B(\M)$. Then, for any
$\mathfrak{M} \in \B$ define the following seminorms on $\M$:
\begin{align}
&p_\mathfrak{M} (X):= \sup_{\xi, \eta \in \mathfrak{M}} |(X \xi| \eta)|, \quad X \in \M  \\
&p^\mathfrak{M} (X):= \sup_{\xi\in \mathfrak{M}} \|X \xi \|, \quad X \in \M  \\
&p^\mathfrak{M}_\dag (X) := \sup_{\xi \in \mathfrak{M}} \{ \|X \xi
\|+ \| X^\dag \xi \| \}, \quad X \in \M.
\end{align}
We call the corresponding locally convex topologies on $\M$ defined
by the families (4.1), (4.2) and (4.3) of seminorms,
\textit{$\B$-uniform topology}, \textit{strongly $\B$-uniform
topology}, resp. \textit{strongly$^*$ $\B$-uniform topology on $\M$}
and denote them by $\tau_u(\B)$, $\tau^u(\B)$, resp. $\tau^u_*(\B)$.
In particular, the $\B(\M)$-uniform topology, the strongly
$\B(\M)$-uniform topology, resp. the strongly$^*$ $\B(\M)$-uniform
topology will be simply called \textit{$\M$-uniform topology},
strongly \textit{$\M$-uniform topology}, resp. \textit{strongly$^*$
$\M$-uniform topology} and will be denoted by $\tau_u(\M)$,
$\tau^u(\M)$, resp. $\tau^u_*(\M)$. In the book of Schm\"{u}dgen
\cite{schmudgen}, these topologies are called \emph{bounded
topologies} and $\tau_u(\B), \ \tau^u(\B)$ are denoted by
$\tau_{\B}, \ \tau^{\B}$, while $\tau_u(\M), \ \tau^u(\M)$ are
denoted by $\tau _{\D}, \ \tau^{\D}$, respectively. The
$\B_f(\D)$-uniform topology, the strongly $\B_f(\D)$-uniform
topology, resp. the strongly$^*$ $\B_f(\D)$-uniform topology is
called \textit{weak topology}, \textit{strong topology}, resp.
\textit{strong$^*$-topology on $\M$}, denoted resp. by $\tau_w$,
$\tau_s$ and $\tau_{s^*}$. All these topologies are related in the
following way:
\begin{equation}
\begin{array}{ccccccccc}
\tau_w & \preceq & \tau_u(\B) & \preceq & \tau_u(\M) \\
\rotatebox{90}{$\succeq$}&
&\rotatebox{90}{$\succeq$}&&\rotatebox{90}{$\succeq$}\\
\tau_s & \preceq & \tau^u(\B) & \preceq & \tau^u(\M) \\
\rotatebox{90}{$\succeq$}&&\rotatebox{90}{$\succeq$}&&\rotatebox{90}{$\succeq$}\\
\tau_{s^*} & \preceq & \tau^u_*(\B) & \preceq & \tau^u_*(\M).
\end{array}
\end{equation}

We investigate now whether $\widetilde{\M_0}[\tau_u(\B)]$ and
$\widetilde{\M_0}[\tau^u_*(\B)]$ are locally convex quasi
$C^*$-algebras over $\M_0$. So, we must check the properties
(T$_1$)-(T$_4$) (stated before and after Definition 3.1) for the
locally convex topologies $\tau_u(\B), \tau^u_*(\B)$ and the
operator $C^*$-norm $\| \cdot \|_0$ on $\M_0$.

(T$_1$) This follows easily for both topologies, since  $\B$ is
admissible and $\M_0 \D \subset \D$.

(T$_2$) Notice that for all $X \in \M_0$ and $\mathfrak{M} \in \B$
we have:
\[
p^\mathfrak{M}_\dag(X)= \sup_{\xi \in \mathfrak{M}} \{ \| X \xi \|+ \|X^\dag \xi\| \}
\leq (2 \sup_{\xi \in \mathfrak{M}} {\| \xi \|}) \|X\|_0,
\]
so by (4.4) we conclude that $\tau_u(\B) \preceq \tau^u_*(\B)
\preceq \| \cdot \|_0$.

(T$_3$) Concerning $\tau_*^u(\B)$, the property (T$_3$) follows
easily from the very definitions. Now, notice the following: For any
$X,Y \in \M_0$ with $XY= YX$ and $Y^*=Y$, one concludes that
\begin{equation}
p_\mathfrak{M}(XY) \leq \| X \|_0 \sup_{\xi \in \mathfrak{M}}
 \left( |Y| \xi \big| \xi \right) , \quad \forall \ \mathfrak{M} \in \B,
\end{equation}
where $|Y|:= (Y^2)^{1/2}$. Then, it follows that for any $X,Y \in \M_0$ with $XY=YX$ and $Y \geq 0$, one has
\[
p_\mathfrak{M} (XY) \leq \|X\|_0 \sup_{\xi \in \mathfrak{M}} (Y \xi
\big|\xi), \quad \forall \ \mathfrak{M}\in \B.
\]

We prove (4.5). From the polar decomposition of $Y$, there is a unique partial
isometry $V$ from $\H$ to $\H$ such that
\[
Y= V|Y|=|Y|V, \quad \ker(V)= \ker (Y) \text{ and } VY= |Y|.
\]
By continuous functional calculus it follows that: $X$ commutes with both $|Y|$
and $|Y|^{1/2}$, but also $V|Y|^{1/2} = |Y|^{1/2}V$. Thus,
\begin{align*}
p_\mathfrak{M}(XY) &= \sup_{\xi,\eta \in
\mathfrak{M}}\bigl|\bigl(XY\xi \big|\eta \bigr)\bigr|
= \sup_{\xi,\eta \in \mathfrak{M}} \bigl| \bigl( V|Y| X \xi \big| \eta \bigr) \bigr| \\
&= \sup_{\xi,\eta \in \mathfrak{M}} \bigl| \bigl( X|Y|^{1/2}\xi
\big| |Y|^{1/2} V \eta \bigr) \bigr|
\leq \sup_{\xi,\eta \in \mathfrak{M}} \|X \|_0 \bigl\| |Y|^{1/2} \xi \bigr\| \bigl\| |Y|^{1/2} \eta \bigr\| \\
& \leq \frac{1}{2} \|X\|_0 \sup_{\xi,\eta \in \mathfrak{M}} \bigl( \bigl\| |Y|^{1/2}
\xi \bigr\|^2 + \bigl\| |Y|^{1/2} \eta \bigr\|^2 \bigr) \\
&\leq \|X \|_0 \sup_{\xi \in \mathfrak{M}} \bigl( |Y| \xi \big|\xi
\bigr), \quad \forall \ \mathfrak{M} \in \B.
\end{align*}
But, we can not say whether (T$_3$) holds for $\tau_u(\B)$. In the
case when $\M_0$ is a von Neumann algebra we have the following:

$\bullet$ \ \ If $\M_0$ is commutative, then (T$_3$) holds for the topology $\tau_w$.

$\bullet$ \ \ If $\M$ is a commutative $O^*$-algebra (see
\cite[p.8]{inoue}) on $\D$ in $\H$, contain\-ing $\M_0$, then
(T$_3$) holds for the topology $\tau_u(\M)$.

Indeed: Suppose that $\M$ is commutative with $\M_0 \subset \M$. For each
$\mathfrak{M} \in \B(\M)$ consider the set
\[
\mathfrak{M}':= \cup \{V \mathfrak{M} : V \text{ partial isometry in } \M_0 \}.
\]
Commutativity of $\M$ implies that $\mathfrak{M}' \in \B(\M)$.
Moreover, $\mathfrak{M} \subset \mathfrak{M}'$. Let now $X,Y \in
\M_0$. Let $Y=V|Y|$ be the polar decomposition of $Y$. Since $\M_0$
is a von Neumann algebra, we have $V \in \M_0$, which implies that
\begin{align*}
p_\mathfrak{M}(XY) &= \sup_{\xi,\eta \in \mathfrak{M}} \bigl|(XY \xi
\big|\eta)\bigr|
= \sup_{\xi,\eta \in \mathfrak{M}} \bigl| \bigl( VX|Y|^{1/2} \xi \big| |Y|^{1/2}\eta \bigr) \bigr| \\
&\leq \|VX\|_0 \sup_{\xi,\eta \in \mathfrak{M}} \bigl\| |Y|^{1/2} \xi \bigr\| \bigl\| |Y|^{1/2} \eta \bigr\| \\
&= \|X\|_0 \sup_{\xi \in \mathfrak{M}} \left( |Y| \xi \big| \xi \right) \\
&= \| X \|_0 \sup_{\xi \in \mathfrak{M}} (Y \xi \big| V^* \xi) \\
& \leq \| X \|_0 \sup_{\xi,\eta \in \mathfrak{M}'} |(Y \xi \big|
\eta) |= \|X\|_0 p_{\mathfrak{M}'} (Y).
\end{align*}

Hence, (T$_3$) holds for $\tau_u(\M)$.

(T$_4$) This property holds for all topologies in (4.4). It suffices
to prove (T$_4$) for the topology $\tau_w$. So, let $X \in
{\overline{\mathcal{U}(\M_0)}}^{\tau_w}$ be arbitrary. Then, there
is a net $\{ X_\alpha\}$ in $\mathcal{U}(\M_0)$ with $X_\alpha
\xrightarrow[\tau_w]{} X$. Notice that the sesquilinear form defined
on $\D \times \D$ by
\[
\D \times \D \ni (\xi,\eta) \mapsto \lim_\alpha (X_\alpha \xi
\big|\eta) \in \mathbb{C},
\]
is bounded.
Hence, $X$ can be regarded as a bounded linear operator on $\H$ such that
\[
\|X\|_0 = 1 \text{ and } (X \xi \big|\eta) = \lim_\alpha (X_\alpha
\xi \big|\eta), \quad \forall \ \xi,\eta \in \D.
\]
Since $\D$ is dense in $\H$, an easy computation shows that
\begin{equation}
(Xx \big|y) =  \lim_\alpha (X_\alpha x \big|y), \quad \forall \ x,y
\in \H.
\end{equation}
This proves that $X \in \M_0 \cap \B(\H)_1 = \mathcal{U}(\M_0)$,
which means that $\mathcal{U}(\M_0)$ is $\tau_w$-closed. A
consequence of (4.6) is now that $\mathcal{U}(\M_0)_+$ is weakly
closed. Similarly we can show that $\widetilde{\M_0}[\tau_w]_{q+}
\cap \M_0 = (\M_0)_+$, therefore (T$_4$) holds for the topology
$\tau_w$ on $\M_0$. From (4.4), (T$_4$) also holds for the
topologies $\tau_u(\B)$ and $\tau^u_*(\B)$.

\medskip

From the preceding discussion we conclude the following

\begin{proposition}
Let $\B$ be  an admissible subset of $\B(\M)$. Then,
$\widetilde{\M_0} [\tau^u_*(\B)]$ and $\widetilde{\M_0}[\tau_{s^*}]$
are locally convex quasi $C^*$-algebras over $\M_0$. If $\M_0$ is a
von Neumann algebra and there is a commutative $O^*$-algebra $\M$ on
$\D$ in $\H$, containing $\M_0$, then $\widetilde{\M_0}[\tau_w]$ and
$\widetilde{\M_0}[\tau_u(\M)]$ are commutative locally convex quasi
$C^*$-algebras over $\M_0$.
\end{proposition}

\begin{remark}{\rm
(1) In general, we do not know whether $\widetilde{\M_0} [\tau_u(\B)]$ and
$\widetilde{\M_0}[\tau_w]$ are locally convex quasi $C^*$-algebras.

(2) The locally convex quasi $C^*$-algebra $\widetilde{\M_0}[\tau_{s^*}]$
over $\M_0$, equals to the completion $\widetilde{\M''_0}[\tau_{s^*}]$ of the
von Neumann algebra $\M''_0$ with respect to the topology $\tau_{s^*}$,
but $\widetilde{\M''_0}[\tau_{s^*}]$ is not necessarily a locally convex quasi
$C^*$-algebra over $\M''_0$, since in general, $\M''_0 \D \not \subset \D$.
In the case when $\M''_0 \D \subset \D$, one has the equality
\[
\widetilde{\M''_0}[\tau_{s^*}] = \widetilde{\M_0} [\tau_{s^*}],
\]
set-theoretically; but, the corresponding locally convex quasi $C^*$-algebras over
$\M_0$ do not coincide. In particular, one has that
\[
\widetilde{\M_0}[\tau_{s^*}]_{cq+} \subsetneq
\widetilde{\M''_0}[\tau_{s^*}]_{cq+}.
\]}
\end{remark}

We present now some properties of the locally convex quasi $C^*$-algebra $\widetilde{\M_0}[\tau_{s^*}]$.

\begin{proposition}
Let $A \in \widetilde{\M_0}[\tau_{s^*}]_{q+}$. Consider the
following:

\emph{(i)} $A \in \widetilde{\M_0}[\tau_{s^*}]_{cq+}$.

\emph{(ii)} $(I+A)^{-1}$ exists and belongs to
$\mathcal{U}(\M_0)_+$.

\emph{(iii)} The closure $\overline{A}$ of $A$ is a positive self-adjoint operator. \\
Then, one has that \emph{(i) $\Rightarrow$ (ii) $\Rightarrow$
(iii)}.
\end{proposition}
\begin{proof}
(i) $\Rightarrow$ (ii) It follows from Proposition 3.2, (1).

(ii) $\Rightarrow$ (iii) Since $(I+A)^{-1}$ is a bounded
self-adjoint operator and $\newline$ $(I+A)^{-1}\D \subset \D$, it follows that
\[
\bigl((I+A)^{-1}(I + A^*)\xi \big|\eta \bigr)= \bigl((I+A^*)\xi
\big|(I+A)^{-1}\eta \bigr) = (\xi \big|\eta),
\]
for all $\xi \in \D(A^*)$ and $\eta \in \D$, which implies
\begin{align*}
(A^* \xi \big|\zeta) &= \bigl((I+A^*)\xi \big|(I+A)^{-1}(I+A^*)\zeta
\bigr) - (\xi \big|\zeta)
\\ &= \bigl(\xi \big|(I+A^*)\zeta \bigr) - (\xi \big|\zeta) = (\xi \big|A^*\zeta), \ \forall
\ \xi, \zeta \in \D(A^*).
\end{align*}
Hence, $\xi \in \D(\overline{A})$ and $\overline{A}\xi = A^* \xi$.
It is now easily seen that $\overline{A}$ is a positive self-adjoint
operator.
\end{proof}

\begin{corollary}
Suppose that $A \in \widetilde{\M''_0}[\tau_{s^*}]$ and $\M''_0 \D \subset \D$.
Then, the following statements are equivalent:

\emph{(i)} $A \in \widetilde{\M''_0}[\tau_{s^*}]_{cq+}$.

\emph{(ii)} $(I+A)^{-1} \in \mathcal{U} (\M''_0)_+$.

\emph{(iii)} $\overline{A}$ is a positive self-adjoint operator.
\end{corollary}
\begin{proof}
From Proposition 4.3 we have that (i) $\Rightarrow$ (ii) $\Rightarrow$ (iii).

(iii) $\Rightarrow$ (i) This follows easily by considering the spectral decomposition of $\overline{A}$.
\end{proof}

It is natural now to ask whether there exists an extended $C^*$-algebra
(abbreviated to $EC^*$-algebra) $\M$ on $\D$ such that
\[
\M_0 \subset \M \subset \widetilde{\M_0}[\tau_{s^*}].
\]

If $\M$ is a closed $O^*$-algebra on $\D$ in $\H$, let $\M_b:= \{ X
\in \M: \overline{X} \in \B(\H) \}$ be the bounded part of $\M$,
where $\B(\H)$ is the $C^*$-algebra of all bounded linear operators
on $\H$. Then, when $\overline{\M_b} \equiv \{ \overline{X}: X \in
\M_b \}$ is a $C^*$-algebra on $\H$ and $(I+X^*X)^{-1} \in \M_b$,
for each $X\in \M$, $\M$ is said to be an \textit{$EC^*$-algebra} on
$\D$.

In this regard, we have the following, which gives a
characterization of certain $EC^*$-algebras on $\D$, through the set
of commutatively quasi-positive elements of
$\widetilde{\M_0}[\tau_{s^*}]$.

\begin{proposition}
Let $\M$ be a closed $O^*$-algebra on $\D$ such that $\M_0 \subset
\M \subset \widetilde{\M_0}[\tau_{s^*}]$ and $\M_b= \M_0$. Then,
$\M$ is an $EC^*$-algebra on $\D$ if and only if $\M_+ \subset
\widetilde{\M_0}[\tau_{s^*}]_{cq+}$.
\end{proposition}
\begin{proof}
Suppose that $\M$ is an $EC^*$-algebra on $\D$ and let $A \in \M_+$
be arbitrary. Then, since $\M_b= \M_0$, $\overline{A}$ is a bounded
positive self-adjoint operator with $(I+\overline{A})^{-1} \in
\mathcal{U}(\M_0)_+$. But, $\widetilde{\M_0}[\tau_{s^*}]$ is a
locally convex quasi $C^*$-algebra (Proposition 4.1), therefore
$\mathcal{U}(\M_0)_+$ is $\tau_{s^*}$-closed. Note that for each $n
\in \mathbb{N}$, the elements $X_n:= \overline{A}(I+\frac{1}{n}
\overline{A})^{-1}$ belong to $(\M_0)_+$, are commuting and $X_n
\xrightarrow[\tau_{s^*}]{} A$, so Definition 3.1 implies that $A \in
\widetilde{\M_0}[\tau_{s^*}]_{cq+}$.

Conversely, suppose that $\M_+ \subset
\widetilde{\M_0}[\tau_{s^*}]_{cq+}$. So, $A \in \M$ implies $A^\dag
A \in \widetilde{\M_0}[\tau_{s^*}]_{cq+}$, therefore $(I+A^\dag
A)^{-1} \in \mathcal{U}(\M_0)_+$ from Proposition 3.2, (1). Now,
since $\M_b = \M_0$ we finally get that $\M$ is an $EC^*$-algebra on
$\D$.
\end{proof}

\section{Structure of commutative locally convex quasi $C^*$-algebras}
Throughout this Section $\A[\tau]$ is a commutative locally convex
quasi $C^*$-algebra over a unital $C^*$-algebra $\A_0$. If the
multiplication of $\A_0$ with respect to the topology $\tau$ is
jointly continuous, then $\A[\tau]$ is a commutative $GB^*$-algebra
\cite[Theorem 2.1]{maria-i-kue}, and so $\A[\tau]$ is isomorphic to
a $*$-algebra of $\mathbb{C}^*$-valued continuous functions on a
compact space, which take the value $\infty$ on at most a nowhere
dense subset \cite[Theorem 3.9]{allan2}, where $\mathbb{C}^*$ is the
extended complex plane in its usual topology as the one-point
compactification of $\mathbb{C}$. The purpose of this Section is to
consider a generalization of the above result in the case when the
multiplication of $\A[\tau]$ is not jointly continuous. As $a^*a$ is
not necessarily defined for $a \in \A[\tau]$, it is impossible to
extend any nonzero multiplicative linear functional $\varphi$  on
$\A_0$ to $\A[\tau]$, like in the case of \cite[Proposition
6.8]{allan1}. Here we show that $\varphi$ is extendable to a
$\mathbb{C}^*$-valued partial multiplicative linear functional
$\varphi'$ on $\A[\tau]_{q+}$, and that $\A[\tau]_{q+}$ is
isomorphic to a wedge of $\mathbb{C}^*$-valued positive functions on
a compact space, which take the value $\infty$ on at most a nowhere
dense subset. This result will be applied in Section 6 for studying
a functional calculus for quasi-positive elements.
 Using the notation given after Definition
3.1, define now a wedge of $\A[\tau]$ as follows:
\[
\A[\tau]_{q+} := \A[\tau] \cap \widetilde{\A_0}[\tau]_{q+} =
\A[\tau] \cap {\overline{(\A_0)_+}}^\tau.
\]
Then, let
\[
\mathfrak{M}(\A_0, \A[\tau]_{q+}):= \{ ax+y : a \in \A[\tau]_{q+}, \
x,y \in \A_0 \},
\]
and denote by $\M(\A_0)$ the Gel'fand space of $\A_0$, i.e. the set
of all nonzero multiplicative linear functionals on $\A_0$, endowed
with the weak$^*$-topology $\sigma(\M(\A_0), \A_0)$. Now, let $a\in
\A[\tau]_{q+}$ and $x,y \in \A_0$. Suppose $x$ is hermitian. Then,
by continuous functional calculus, $x$ is uniquely decomposed in the
following way:
\begin{align*}
&x=x_+-x_-, \ \ x_+, \ \ x_- \in (\A_0)_+, \ \ x_+ x_- =0 \\
&|x| \equiv (x^*x)^{1/2}=x_++x_- \in (\A_0)_+.
\end{align*}
Hence, $a|x|, \ ax_+, \ ax_- \in \A[\tau]_{q+}$, and by (1) and (2)
of Proposition 3.2, $(\mathit{1} +a|x|)^{-1}$, $a|x|(\mathit{1}+
a|x|)^{-1} \in (\A_0)_+$. Furthermore, since
\[
a|x| (\mathit{1} +a|x|)^{-1}-ax_+(\mathit{1}+a|x|)^{-1} =ax_-(\mathit{1}+ a|x|)^{-1}
 \in \widetilde{\A_0}[\tau]_{q+},
\]
Proposition 3.2,(4) implies that $ax_+ (\mathit{1}+ a|x|)^{-1} \in
(\A_0)_+$. Similarly, $ax_-(\mathit{1}+ a|x|)^{-1} \in (\A_0)_+$.
Hence, we have
\begin{align*}
(ax+y)(\mathit{1}+ a|x|)^{-1} = ax_+ (\mathit{1} + a|x|)^{-1} &-ax_-
(\mathit{1}+ a|x|)^{-1} \\ &+ y(\mathit{1}+ a|x|)^{-1} \in \A_0.
\end{align*}
Since a general element $x$ of $\A_0$ is a linear combination of two
hermitian elements of $\A_0$, we finally obtain that
\begin{align*}
(ax+y) (\mathit{1}+ a|x|)^{-1} \in \A_0, \ \forall \ a \in
\A[\tau]_{q+} \ \text {and } \ x,y \in \A_0.
\end{align*}
Indeed: Let $x$ be arbitrary in $\A_0$. Then, $x=x_1+ix_2$, with
$x_1$ and $x_2$ hermitian. An easy computation shows that
\begin{align*}
|x|  & \leq |x_1|+|x_2|, \ |x_j| \le |x|,
\ (1+a|x_j|)(1+a|x|)^{-1} \in \widetilde{\A_0}[\tau]_{q+}, \\
& \text{ and } \ 1- (1+a|x_j|)(1+a|x|)^{-1} \in
\widetilde{\A_0}[\tau]_{q+}, \ j=1,2.
\end{align*}
The latter together with Proposition 3.2,(4) gives
$(1+a|x_j|)(1+a|x|)^{-1} \in (\A_0)_+$; moreover, from the above
$(ax_j+y)(1+a|x_j|)^{-1} \in \A_0$. Thus, for $j=1,2$, we get
\[
(ax_j+y)(1+a|x|)^{-1} =
((ax_j+y)(1+a|x_j|)^{-1})((1+a|x_j|)(1+a|x|)^{- 1}) \in \A_0,
\]
which implies
\[
(ax+y)(1+a|x|)^{-1}=(ax_1+y)(1+a|x|)^{-1}+ix_2(1+a|x|)^{-1} \in\A_0.
\]
Hence, the elements $\varphi((\mathit{1}+a|x|)^{-1})$ and
$\varphi((ax+y)(\mathit{1}+a|x|)^{-1})$ are complex numbers for each
$\varphi \in \M(\A_0)$, so that we can consider the correspondence
\begin{align*}
&\varphi': \mathfrak{M}( \A_0, \A[\tau]_{q+}) \longrightarrow
\mathbb{C}^* \equiv \mathbb{C}
\cup \{ \infty \}, \text{ with } \\
&\qquad ax + y \mapsto \varphi'(ax+y) = \begin{cases}
\frac{\varphi((ax+y) (\mathit{1}+a|x|)^{-1})}{\varphi((\mathit{1}+a|x|)^{-1})} &
\text{ if } \varphi((\mathit{1}+a|x|)^{-1}) \neq 0 \\
\infty & \text{ if } \varphi((\mathit{1}+a|x|)^{-1}) =0.
\end{cases}
\end{align*}
Then, we have

\begin{lemma}
The following statements hold:

\emph{(1)} For every $\varphi \in \M(\A_0)$ the correspondence $\varphi'$, given above, is well-defined.

\emph{(2)} Let $a \in \A[\tau]_{q+}$ and $x \in \A_0$. Then,
$(\mathit{1}+a)^{-1}$ exists in $\A_0$ \emph{(from Proposition
3.2,(1))} and we have:

\emph{(i)} $\varphi((\mathit{1}+a|x|)^{-1})=0$ implies $\varphi((\mathit{1}+a)^{-1})=0, \varphi \in \M(\A_0)$.

\emph{(ii)} $\varphi((\mathit{1}+a)^{-1})=0$ and $\varphi(x) \neq 0$ imply $\varphi((\mathit{1}+a|x|)^{-1})=0,
\varphi \in \M(\A_0)$.
\end{lemma}
\begin{proof}
(1) Let $a,b \in \A[\tau]_{q+}$ and $x,y,z,w \in \A_0$ such that
$ax+y=bz+w$. Then, for every $\varphi \in \M(\A_0)$ one has that
\begin{equation}
\varphi((\mathit{1}+a|x|)^{-1})=0 \Leftrightarrow \varphi((\mathit{1}+b|z|)^{-1})=0.
\end{equation}
Indeed: We first show (5.1) in case $x$ and $z$ are hermitian. Since $ax+y= bz+w$, we have
\[
(\mathit{1}+ a|x|)-2ax_-+y = (\mathit{1}+b|z|)-2bz_-+w.
\]
We multiply the last equality by $ (\mathit{1}+
a|x|)^{-1}(\mathit{1}+b|z|)^{-1}$ and get
\begin{align*}
&(\mathit{1}+b|z|)^{-1}- 2ax_- (\mathit{1}+ a|x|)^{-1} (\mathit{1}+b|z|)^{-1}+y (\mathit{1}+ a|x|)^{-1}
(\mathit{1}+b|z|)^{-1} \\
& \quad = (\mathit{1}+a|x|)^{-1} - 2bz_- (\mathit{1}+b|z|)^{-1} (\mathit{1}+ a|x|)^{-1}+w
(\mathit{1}+ a|x|)^{-1} (\mathit{1}+b|z|)^{-1}.
\end{align*}
This implies that for every $\varphi \in \M(\A_0)$
\begin{equation}
\varphi((\mathit{1}+ a|x|)^{-1})=0 \Leftrightarrow \varphi((\mathit{1}+b|z|)^{-1}) =0.
\end{equation}
We next prove (5.1) in the case when $x$ and $z$ are arbitrary
elements of $\A_0$. Then, the elements $x,y,z$ and $w$ are
decomposed into
\begin{align*}
x= x_1+ix_2, \quad y=y_1+iy_2, \quad
z= z_1+iz_2, \quad w=w_1+iw_2,
\end{align*}
where $x_j,y_j,z_j,w_j \ (j=1,2)$ are hermitian elements in $\A_0$
that satisfy the equations:
\begin{equation}
ax_1+y_1= bz_1+w_1, \quad ax_2 +y_2 = bz_2 + w_2.
\end{equation}
We show now that
\begin{equation}
\begin{array}{ll}
\varphi((\mathit{1}+ a|x|)^{-1})=0 \ \ \Leftrightarrow &\text{ either } \ \varphi((\mathit{1}+ a|x_1|)^{-1})=0 \\
&\text{ or } \ \varphi((\mathit{1}+ a|x_2|)^{-1})=0.
\end{array}
\end{equation}
Suppose that $\varphi((\mathit{1}+ a|x_1|)^{-1}) \neq 0 $ and
$\varphi((\mathit{1}+ a|x_2|)^{-1}) \neq 0$. Then,
\begin{align*}
&(\mathit{1}+ a(|x_1|+|x_2|))^{-1}-(\mathit{1}+ a|x_1|)^{-1} (\mathit{1}+ a|x_2|)^{-1} \\
& \quad = (\mathit{1}+a(|x_1|+|x_2|))^{-1}
(a|x_1|(\mathit{1}+ a|x_1|)^{-1})(a|x_2|(\mathit{1}+ a|x_2|)^{-1})
\in (\A_0)_+,
\end{align*}
whence
\[
\varphi((\mathit{1}+ a(|x_1|+|x_2|))^{-1}) \geq \varphi((\mathit{1}+ a|x_1|)^{-1}(\mathit{1}+ a|x_2|)^{-1})>0.
\]
Furthermore, since $|x| \leq |x_1| + |x_2|$, we have
\[
0 < \varphi((\mathit{1}+a(|x_1|+|x_2|))^{-1}) \leq \varphi((\mathit{1}+ a|x|)^{-1}).
\]
Hence, $\varphi((\mathit{1}+ a|x|)^{-1}) \neq 0$. Conversely,
suppose $\varphi((\mathit{1}+ a|x_1|)^{-1})=0$ or
$\varphi((\mathit{1}+ a|x_2|)^{-1})=0$. Then, since $(\mathit{1}+
a|x_j|)^{-1}\geq (\mathit{1}+ a|x|)^{-1}$, $j=1,2$, we have that
$\varphi((\mathit{1}+ a|x|)^{-1})=0$.

Now from (5.2), (5.3) and (5.4) we get
\begin{align*}
\varphi((\mathit{1}+ a|x|)^{-1})=0
&\Leftrightarrow \varphi((\mathit{1}+ a|x_1|)^{-1})=0 \text{ or } \varphi((\mathit{1}+ a|x_2|)^{-1})=0 \\
&\Leftrightarrow \varphi((\mathit{1}+ b|z_1|)^{-1})=0 \text{ or } \varphi((\mathit{1}+b|z_2|)^{-1})=0 \\
&\Leftrightarrow \varphi((\mathit{1}+b|z|)^{-1})=0.
\end{align*}
Thus, (5.1) has been shown.
Now, by assumption $ax+y= bz+w$, consequently
\[
\varphi'(ax+y)= \infty \Leftrightarrow \varphi'(bz+w)=\infty.
\]
On the other hand, from (5.1) it follows that
\[
\varphi'(ax+y) < \infty \Leftrightarrow \varphi'(bz+w)<\infty.
\]
In this case,
\[
\varphi'(ax+y)= \frac{\varphi((ax+y)(\mathit{1}+a|x|)^{-1} (\mathit{1}+b|z|)^{-1})}
{\varphi((\mathit{1}+a|x|)^{-1}) \varphi((\mathit{1}+b|z|)^{-1})} = \varphi'(bz+w)
\]
and this completes the proof of (1).

(2) (i) Suppose $\varphi((\mathit{1}+a|x|)^{-1})=0, \varphi \in \M(\A_0)$, Then,
\begin{align*}
(\mathit{1}+a)^{-1}
&= (\mathit{1}+a|x|)^{-1} (\mathit{1}+a|x|) (\mathit{1}+a)^{-1} \\
&= (\mathit{1}+a|x|)^{-1} ((\mathit{1}+a)^{-1}+ |x| a (\mathit{1}+a)^{-1}) \\
&= (\mathit{1}+a|x|)^{-1} ((\mathit{1}+a)^{-1}+|x| -|x|(\mathit{1}+a)^{-1}) \\
&= (\mathit{1}+a|x|)^{-1} ((\mathit{1}-|x|)(\mathit{1}+a)^{-1}+|x|),
\end{align*}
where $(\mathit{1}-|x|) (\mathit{1}+a)^{-1} + |x| \in \A_0$.
So applying $\varphi$ we have $\varphi((\mathit{1}+a)^{-1})=0$.

(ii) Suppose that $\varphi((\mathit{1}+a)^{-1} )=0$  and $\varphi(x)
\neq 0$, $\varphi \in \M(\A_0)$. Then, we apply $\varphi$ to the
final result of the preceding calculation in (i) and we take
\[
\varphi((\mathit{1}+a|x|)^{-1} ) \varphi(|x|)=0.
\]
Since $\varphi(x) \neq 0$ if and only if $\varphi(|x|) \neq 0$,
clearly we have $\varphi((\mathit{1}+a|x|)^{-1} )=0$.
\end{proof}

\begin{proposition}
For $\varphi \in \M(\A_0)$, the well defined map $\varphi'$ has the following properties:

\emph{(1)} $\varphi' \supset \varphi$ (i.e., $\varphi'$ is an extension of $\varphi$);

\emph{(2)} $\varphi'(ax+y) = \varphi'(a) \varphi(x) + \varphi(y)$
and $\varphi'(ax) = \varphi'(a) \varphi(x)$, whenever $a \in
\A[\tau]_{q+}$ and $x,y \in \A_0$ such that $\varphi'(a) \varphi(x)
\neq \infty \cdot 0$;

\emph{(3)} $\varphi'(a+b) = \varphi'(a)+ \varphi'(b)$, for all $a,b
\in \A[\tau]_{q+}$;

\emph{(4)} $\varphi'(\lambda a) = \lambda \varphi'(a)$, for all
$\lambda \in \mathbb{C}$ and $a \in \A[\tau]_{q+}$, where $0 \cdot
\infty = 0$.
\end{proposition}
\begin{proof}
(1) It is trivial.

(2) Suppose that $\varphi'(a) \varphi(x) \neq \infty \cdot 0$,
$\varphi \in \M(\A_0)$. Then, from the definition of $\varphi'$ and
Lemma 5.1,(2), we have the following implications (considering
separately the cases where $\varphi'(a)$ is infinite or not):

$\bullet$
\begin{tabular}[t]{ccc}
$\varphi'(ax+y) = \infty $ \qquad &$\Leftrightarrow$ &$\varphi'(a)=\infty$ \\
$\Updownarrow$ & &$\Updownarrow$ \\
$\varphi((\mathit{1}+a|x|)^{-1} )=0$ & &$ \varphi((\mathit{1}+a)^{-1} )=0$ \\
& &$\Downarrow$ \\
& &$\varphi'(a) \varphi(x) + \varphi(y) =\infty$.
\end{tabular}

\medskip

$\bullet$
\begin{tabular}[t]{ccc}
$\varphi'(ax+y) < \infty $ \qquad &$\Leftrightarrow$ &$\varphi'(a)<\infty$ \\
$\Updownarrow$ & &$\Updownarrow$ \\
$\varphi((\mathit{1}+a|x|)^{-1} )\neq 0$ & &$ \varphi((\mathit{1}+a)^{-1} ) \neq 0$.
\end{tabular}

\medskip

\noindent
So, in this case we also get
\begin{align*}
\varphi'(ax+y)&= \frac{\varphi(ax(\mathit{1}+a|x|)^{-1})}{\varphi((\mathit{1}+a|x|)^{-1})} + \varphi(y) \\
&= \frac{\varphi(a(\mathit{1}+a)^{-1})\varphi(x)}{\varphi((\mathit{1}+a)^{-1})} + \varphi(y)
= \varphi'(a) \varphi(x) + \varphi(y),
\end{align*}
and this completes the proof of (2).

(3) Observe that for any $a,b \in \A[\tau]_{q+}$, one has
\begin{align*}
(\mathit{1}+a)^{-1} (\mathit{1}+b)^{-1} = &(\mathit{1}+a+b)^{-1} ((\mathit{1}+a)^{-1}  (\mathit{1}+b)^{-1} \\
&+ a (\mathit{1}+a)^{-1} (\mathit{1}+b)^{-1}
+ (\mathit{1}+a)^{-1} b ( \mathit{1}+b)^{-1}),
\end{align*}
where $(\mathit{1}+a)^{-1} (\mathit{1}+b)^{-1}+ a(\mathit{1}+a)^{-1}
(\mathit{1}+b)^{-1} + (\mathit{1}+a)^{-1} b(\mathit{1}+b)^{-1} \in
\A_0$ (see Proposition 3.2). Thus, applying any $\varphi\in
\M(\A_0)$ to the last equality we conclude that
\begin{equation}
\begin{array}{ll}
&\varphi((\mathit{1}+a+b)^{-1})=0 \text{ implies either } \\
&\varphi((\mathit{1}+a)^{-1})=0 \text{ or } \varphi((\mathit{1}+b)^{-1})=0.
\end{array}
\end{equation}
Conversely, observe that
\[
(\mathit{1}+a)^{-1}
= (\mathit{1}+a+b)^{-1} + b(\mathit{1}+a+b)^{-1} (\mathit{1}+a)^{-1},
\]
where $b(\mathit{1}+a+b)^{-1} \in (\A_0)_+$ by Proposition 3.2,(4),
since $(a+b) (\mathit{1}+a+b)^{-1}-a(\mathit{1}+a+b)^{-1} =
b(\mathit{1}+a+b)^{-1} \in \widetilde{\A_0}[\tau]_{q+}$ with $(a+b)
(\mathit{1}+a+b)^{-1} \in (\A_0)_+$. So, taking also into account an
analogous equality for $(\mathit{1}+b)^{-1}$, as well as (5.5) we
have that
\begin{align*}
&\varphi((\mathit{1}+a+b)^{-1})=0 \Leftrightarrow \text{ either } \varphi((\mathit{1}+a)^{-1}) =0 \text{ or } \\
&\varphi((\mathit{1}+b)^{-1}) =0, \quad \forall \ \varphi \in
\M(\A_0).
\end{align*}
Using now the preceding equivalence, clearly we conclude that:

$\bullet$ \ \ $\varphi'(a+b) = \infty \Leftrightarrow \text{ either } \varphi'(a)= \infty \text{ or }
\varphi'(b) = \infty$; thus,
\[
\varphi'(a+b) = \varphi'(a) + \varphi'(b) = \infty; \text{ or }
\]

$\bullet$ \ \ $\varphi'(a+b) < \infty \Leftrightarrow \varphi'(a) < \infty \text{ and } \varphi'(b) < \infty$. \\
In this case,
\begin{align*}
& \varphi'(a+b) \\
& \, = \frac{\varphi(a(\mathit{1}+a)^{-1} (\mathit{1}+b)^{-1} (\mathit{1}+a+b)^{-1}+
b (\mathit{1}+a)^{-1}(\mathit{1}+b)^{-1}(\mathit{1}+a+b)^{-1})}{\varphi((\mathit{1}+a)^{-1})
\varphi((\mathit{1}+b)^{-1}) \varphi((\mathit{1}+a+b)^{-1})} \\
& \, = \frac{ \varphi(a(\mathit{1}+a)^{-1})}{\varphi((\mathit{1}+a)^{-1})} +
\frac{\varphi(b(\mathit{1}+b)^{-1})}{\varphi((\mathit{1}+b)^{-1})}\\
& \, = \varphi'(a)+ \varphi'(b).
\end{align*}

(4) It follows from (2) by replacing $x$  with $\lambda \mathit{1}, \lambda \in \mathbb{C}$, and $y$ with $0$.
\end{proof}

\begin{remark}{\rm
In order to have all the values of $\varphi'$ fully determined, we
need to define the following:

$\bullet$ \ \ $\varphi'(a)\varphi(x)$, \ \ $\varphi'(ax) +
\varphi'(bx)$ \ and \ $\varphi'(a)\varphi(x_1) +
\varphi'(a)\varphi(x_2)$, \ for any $a,b \in \A[\tau]_{q+}$ and
$x_1, x_2 \in \A_0$.

From Proposition 5.2 we conclude that:

{\bf (i)} \ \ $\varphi'(a)\varphi(x) = \varphi'(ax)$, \ for any \ $a
\in \A[\tau]_{q+}$ and $x \in \A_0$ \ with \ \
$\varphi'(a)\varphi(x) \neq \infty \cdot 0$.

{\bf (ii)} \ \ $\varphi'(ax) + \varphi'(bx) = \varphi'((a +b)x)$, \
for any \ $a,\;b \in \A[\tau]_{q+}$ and $x \in \A_0$ \ with \ \
either $\varphi'(a)\varphi(x) \neq \infty \cdot 0$ \ or \
$\varphi'(b)\varphi(x) \neq \infty \cdot 0$.

{\bf (iii)} \ \ $\varphi'(a)\varphi(x_1 + x_2) = \varphi'(a(x_1 +
x_2))$, \ for any $a \in \A[\tau]_{q+}$ and $x_1, x_2 \in \A_0$ \
with \ $\varphi'(a(x_1 + x_2)) \neq \infty \cdot 0$.

Furthermore, the definition of $\varphi'$ and Proposition 5.2 imply
that:

{\bf (1)} When $\varphi'(a) = \infty$ and $\varphi(x) = 0$, the
value $\varphi'(ax)$ of $\varphi'$ depends upon $a$ and $x$. For
instance,

$\bullet$ \ \ $x=0 \Rightarrow  \varphi'(ax)= \varphi'(0) = \varphi(0) = 0$;

$\bullet$ \ \ $x=(\mathit{1}+a)^{-1} \Rightarrow \varphi'(a(\mathit{1}+a)^{-1})
= \varphi(a(\mathit{1}+a)^{-1}) = \varphi(\mathit{1}-(\mathit{1}+a)^{-1}) = \mathit{1}$.

{\bf (2)} For $a,b \in \A[\tau]_{q+}$ and $x \in \A_0$ such that
either $\varphi'(a) \varphi(x) = \infty \cdot 0$ or $\varphi'(b)
\varphi(x) = \infty \cdot 0$, the value $\varphi'((a+b)x)$ clearly
depends upon $a,b$ and $x$.

{\bf (3)} For $a \in \A[\tau]_{q+}$ and $x_1, x_2 \in \A_0$ such
that either $\varphi'(a) \varphi(x_1) = \infty \cdot 0$ or
$\varphi'(a) \varphi(x_2)= \infty \cdot 0$, then again the value
$\varphi'(a (x_1+x_2))$ depends upon $a$, $x_1$ and $x_2$.

{\bf Conclusion.} We define the requested values of $\varphi'$ by
(i), (ii) and (iii), for any $a,\;b \in \A[\tau]_{q+}$ and $x_1, x_2
\in \A_0$.}
\end{remark}

\begin{remark}{\rm
We do not know whether $\varphi'$ is defined or not on the linear
span of $\mathfrak{M}(\A_0, \A[\tau]_{q+})$.}
\end{remark}

Now, for any $a \in \A[\tau]_{q+}$ and $x,y \in A_0$, we fix the
notation:
\[
\widehat{ax+y} (\varphi) \equiv  \varphi'(ax+y), \quad \varphi \in \M(\A_0).
\]
Then, we have the following

\begin{proposition}
$\widehat{ax+y}$ is a $\mathbb{C}^*$-valued continuous function on the compact Hausdorff
space $\M(\A_0)$, which takes the value $\infty$ on at most $a$ nowhere dense subset of $\M(\A_0)$.
\end{proposition}
\begin{proof}
We shall show that the set
\[
N_{\widehat{ax+y}} \equiv \{ \varphi \in \M(\A_0): \widehat{ax+y} (\varphi) = \infty \},
\]
is a nowhere dense closed subset of $\M(\A_0)$.
Notice that
\begin{equation}
N_{\widehat{ax+y}} = \{ \varphi \in \M(\A_0) : \varphi((\mathit{1}+a|x)^{-1}) = 0 \},
\end{equation}
from which it follows that $N_{\widehat{ax+y}}$ is closed.
Now, suppose that
\[
\exists \ \mathcal{U} \text{ non-empty open subset of } \M(\A_0) \text{ with }
\mathcal{U} \subset N_{\widehat{ax+y}}.
\]
From the commutative Gel'fand-Naimark theorem, $\A_0 \simeq \mathcal{C}(\M(\A_0))$,
up to an isometric $*$-isomorphism. Thus, using Urysohn's lemma for $\M(\A_0)$ we get that
\[
\exists \ b \in \A_0: \|b\|_0=1 \text{ and } \hat{b}(\varphi) =
\varphi(b)=0, \quad \forall \ \varphi \not \in \mathcal{U}.
\]
But this together with (5.6) and the fact that $\mathcal{U} \subset N_{\widehat{ax+y}}$, implies
\[
\varphi(b(\mathit{1}+a|x|)^{-1})=0, \quad \forall \ \varphi \in
\M(\A_0).
\]
The afore-mentioned identification $\A_0 \simeq \mathcal{C}(\M(\A_0))$ gives
now $b(\mathit{1}+a|x|)^{-1}=0$, which clearly yields $b=0$, a contradiction to $\|b\|_0 =1$.
Hence, $N_{\widehat{ax+y}}$ is a nowhere dense closed subset of $\M(\A_0)$.

Next we show that $\widehat{ax+y}$ is continuous on $\M(\A_0)$. Put
\[
z\equiv (\mathit{1}+a|x|)^{-1} \text{ and } w \equiv ax(\mathit{1}+a|x|)^{-1}.
\]
Take an arbitrary $\varphi_0 \in \M(\A_0)$ and consider the cases:

$\bullet$ \ $\widehat{ax+y} (\varphi_0) \neq \infty$, i.e., $\hat{z}(\varphi_0) \neq 0$. \\
From the continuity of $\hat{z}$ there is a neighborhood $\mathcal{U}_{\varphi_0}$
of $\varphi_0$ with $\hat{z}(\varphi) \neq 0$, for all $\varphi \in \mathcal{U}_{\varphi_0}$.
Thus, we get
\[
\widehat{ax+y} (\varphi)= \frac{\hat{w}(\varphi)}{\hat{z} (\varphi)}
+ \hat{y}(\varphi), \quad \forall \ \varphi \in
\mathcal{U}_{\varphi_0}, \] where all functions
$\hat{w},\hat{z},\hat{y}$ are continuous at $\varphi_0$, so that the
same is true for $\widehat{ax+y}$.

$\bullet$ \ $\widehat{ax+y} (\varphi_0)=\infty$, i.e., $\hat{z}(\varphi_0)=0$.\\
Take an arbitrary net $\{\varphi_\alpha \}$ in $\M(\A_0)$ such that $\varphi_\alpha \rightarrow \varphi_0$,
with respect to the weak$^*$-topology $\sigma(\M(\A_0), \A_0)$.
Then,
\[
\hat{z}(\varphi_\alpha) \rightarrow \hat{z}(\varphi_0)=0,
\]
where $\hat{z}(\varphi_\alpha) \neq 0$, since $N_{\widehat{ax+y}}$ is a nowhere dense subset of $\M(\A_0)$.
Since
\begin{align*}
|\widehat{ax}(\varphi_\alpha)|
&= \dfrac{\varphi_\alpha((ax^* ((\mathit{1}+ a|x|)^{-1})(ax(\mathit{1}+
a|x|)^{-1}))^{1/2}}{\varphi_\alpha((\mathit{1}+ a|x|)^{-1})} \\
&= \dfrac{\varphi_\alpha((a(\mathit{1}+ a|x|)^{-1})(x^*xa(\mathit{1}+
a|x|)^{-1}))^{1/2}}{\varphi_\alpha((\mathit{1}+ a|x|)^{-1})} \\
&= \dfrac{\varphi_\alpha((a|x|(\mathit{1}+ a|x|)^{-1})^2)^{1/2}}{\varphi_\alpha((\mathit{1}+ a|x|)^{-1})} \\
&= \dfrac{\varphi_\alpha(a|x|(\mathit{1}+ a|x|)^{-1})}{\varphi_\alpha((\mathit{1}+ a|x|)^{-1})} \\
&= \dfrac{\varphi_\alpha(\mathit{1}- (\mathit{1}+ a|x|)^{-1})}{\varphi_\alpha((\mathit{1}+ a|x|)^{-1})} \\
&= \dfrac{1}{\widehat{z}(\varphi_\alpha)}-1,
\end{align*}
it follows that $\displaystyle \lim_\alpha
\widehat{ax}(\varphi_\alpha)=\infty$, which implies
\[
\displaystyle \lim_\alpha \widehat{ax+y} (\varphi_\alpha)=\infty =
\widehat{ax+y}(\varphi_0).
\]
This completes the proof of the continuity of $\widehat{ax+y}$ at
$\varphi_0$; so the proof of Proposition 5.5 is finished.
\end{proof}

All the above lead to the following

\begin{definition}{\rm
Let $W$ be a completely regular topological space and $\mathcal{F}(W)_+$ the set of all
$\mathbb{C}^*$-valued positive continuous functions on $W$, which take the value $\infty$
on at most a nowhere dense subset $W_0$ of $W$. Then, $\mathcal{F}(W)_+$ is said to be a
\textit{wedge} on $W$, if for any $f,g \in \mathcal{F}(W)_+$ and $\lambda \geq 0$, the
functions $f+g$ and $\lambda f$ defined pointwise on $W_0$ on which $f$ and $g$ are both
finite, are extendible to $\mathbb{C}^*$-valued positive continuous functions on $W$ that
also belong to $\mathcal{F}(W)_+$. We keep the same symbols $f+g$ and $\lambda f$ for the respective extensions. }
\end{definition}

Consider now the set
\[
\mathcal{F}(W) \equiv \{ fg_0+h_0: f \in \mathcal{F}(W)_+, g_0, h_0 \in \mathcal{C}(W)\},
\]
where $\mathcal{C}(W)$ is the $*$-algebra of all continuous $\mathbb{C}$-valued functions on $W$.
Then, the set $\mathcal{F}(W)$ fulfils the following conditions:

$\bullet$ \ $(f_1+f_2)g_0 = f_1 g_0 + f_2 g_0$,

$\bullet$ \ $(\lambda f) g_0 = \lambda (f g_0)$,

$\bullet$ \ $f(g_0+h_0)= f g_0 +f h_0$,
\\
for all $f, f_1, f_2 \in \mathcal{F}(W)_+$, $g_0,h_0 \in \mathcal{C}(W) $ and $\lambda \geq 0$.

\begin{definition}{\rm
We call $\mathcal{F}(W)$ \textit{the set of $\mathbb{C}^*$-valued positive continuous functions
on $W$ generated by the wedge $\mathcal{F}(W)_+$ and the $*$-algebra} $\mathcal{C}(W)$. }
\end{definition}

In this regard (see also Remark 5.3), we have the following

\begin{theorem}
Let $\mathcal{F}(\M(\A_0))_+ \equiv \{ \hat{a} : a \in \A[\tau]_{q+}
\}$. Then,

\emph{(1)} $\mathcal{F}(\M(\A_0))_+$ is a wedge on $\M(\A_0)$.

\emph{(2)} The map $\Phi: \mathfrak{M}(\A_0, \A[\tau]_{q+})
\rightarrow \mathcal{F}(\M(\A_0)): ax+y \mapsto \widehat{ax+y}$, is
a bijection satisfying the properties:

\emph{(i)} $\Phi(\A[\tau]_{q+}) = \mathcal{F}(\M(\A_0))_+$, with \\
\hspace*{10mm} $\Phi(a+b)= \Phi(a)+ \Phi(b)$ and $\Phi(\lambda a) =
\lambda \Phi(a)$,
for all $a,b \in \A[\tau]_{q+}$ and \\
\hspace*{10mm} $\lambda \geq 0$.

\emph{(ii)} $\Phi(\A_0)= \mathcal{C}(\M(\A_0))$, $\Phi$ being an isometric $*$-isomorphism from $\A_0$\\
\hspace*{11mm} onto $\mathcal{C}(\M(\A_0))$.

\emph{(iii)} $\Phi(ax)= \Phi(a) \Phi(x)$, for all $a \in \A[\tau]_{q+}$ and $x \in \A_0$. \\
\hspace*{12mm} $\Phi((a+b)x)=(\Phi(a)+\Phi(b)) \Phi(x)$, for all $a,b \in \A[\tau]_{q+}$ and $x \in \A_0$. \\
\hspace*{12mm} $\Phi(\lambda ax) = \lambda \Phi(a) \Phi(x)$, for all
$a \in \A[\tau]_{q+}, x \in \A_0$ and $\lambda \geq 0$.
\\
\hspace*{12mm} $\Phi(a(x_1+x_2)) = \Phi(a) ( \Phi(x_1)+\Phi(x_2))$, for all
$a \in \A[\tau]_{q+}$ and $x_1, x_2 \in$ \\
\hspace*{12mm} $\A_0$.
\end{theorem}
\begin{proof}
The statements (1), (2)(i) and (2)(ii) follow from Propositions 5.2
and 5.5. We show the statement (2)(iii). Let $a \in \A[\tau]_{q+}$
and $x \in \A_0$. From Proposition 5.5, $\hat{a}$ and $\widehat{ax}$
are $\mathbb{C}^*$-valued continuous functions on $\M(\A_0)$ that
take the value $\infty$ on at most a nowhere dense subset of
$\M(\A_0)$. Hence, the set
\[
\mathcal{K} \equiv \{ \varphi \in \M(\A_0): \hat{a}(\varphi) < \infty \text{ and } \widehat{ax} (\varphi) < \infty \}
\]
is dense in $\M(\A_0)$ and
\[
\widehat{ax}(\varphi) = \hat{a}(\varphi) \hat{x}(\varphi), \forall \
\varphi \in \mathcal{K},
\]
therefore by the continuity of $\hat{a}$ and $\widehat{ax}$ we
conclude that $\widehat{ax}=\hat{a}\hat{x}$, from which it follows
that $\Phi(ax) = \Phi(a) \Phi(x)$. The rest of the properties in
(2)(iii) are similarly proved.
\end{proof}

\section{Functional calculus for quasi-positive elements}

Throughout this Section $\A[\tau]$ is a commutative locally convex
quasi $C^*$-algebra over a $C^*$-algebra $\A_0$. Here we shall
consider a functional calculus for the quasi-positive elements of
$\A[\tau]$, resulting, for instance, to consideration of the quasi
$n$th-root of an element $a \in \A[\tau]_{q+}$ (see Corollary 6.7).
For this purpose, we first need to extend the multiplication of
$\A[\tau]$.

\begin{definition}{\rm
Let $a,b \in \A[\tau]_{q+}$; $a$ is called \textit{left-multiplier}
of $b$, and we write $a \in L(b)$, if there exist nets $\{ x_\alpha
\}, \{y_\beta \}$ in $(\A_0)_+$ such that $x_\alpha
\xrightarrow[\tau]{} a$, $y_\beta \xrightarrow[\tau]{} b$ and
$x_\alpha y_\beta \xrightarrow[\tau]{}c$ (in the sense that the
double indexed net $\{x_\alpha y_\beta\}$ converges to $c$). The
product of $a,b$ denoted by $ab$ is given as follows
\[
ab: =c= \tau\text{-}\lim_{\alpha,\beta}x_\alpha y_\beta.
\]}
\end{definition}

\begin{lemma}
The product $ab$ is well-defined, in the sense that it is independent of the selection of
the nets $\{x_\alpha\}, \{y_\beta\}$.
\end{lemma}
\begin{proof}
Let $\{x_\alpha\}, \{y_\beta\}$ be two nets in $(\A_0)_+$ such that
\[
x_\alpha \xrightarrow[\tau]{} a, \ y_\beta \xrightarrow[\tau]{} b \text{ and } x_\alpha y_\beta \xrightarrow[\tau]{} c.
\]
Then (also see Proposition 3.2)
\begin{align*}
&(\mathit{1}+x_\alpha)^{-1} x_\alpha y_\beta (\mathit{1}+y_\beta)^{-1} (\mathit{1}+c)^{-1}-(\mathit{1}+a)^{-1}
c (\mathit{1}+c)^{-1} (\mathit{1}+b)^{-1} \\
&= ((\mathit{1}+x_\alpha)^{-1} x_\alpha y_\beta (\mathit{1}+y_\beta)^{-1} (\mathit{1}+c)^{-1} -(\mathit{1}
+x_\alpha)^{-1}c(\mathit{1}+c)^{-1} (\mathit{1}+y_\beta)^{-1}) \\
& \quad + ((\mathit{1} +x_\alpha)^{-1}c (\mathit{1}+c)^{-1}(\mathit{1}+y_\beta)^{-1}-(\mathit{1}+a)^{-1}
c (\mathit{1}+c)^{-1}( \mathit{1}+y_\beta)^{-1}) \\
&\quad + ((\mathit{1}+a)^{-1} c(\mathit{1}+c)^{-1} (\mathit{1}+y_\beta)^{-1} - (\mathit{1}+a)^{-1}
c(\mathit{1}+c)^{-1} (\mathit{1}+b)^{-1}).
\end{align*}
As we have seen in the proof of Proposition 3.2,(1) $(\mathit{1}+x_\alpha)^{-1} \xrightarrow[\tau]{} a$,
so taking $\tau$-limits in the preceding equality, we conclude that
\[
(\mathit{1}+x_\alpha)^{-1}x_\alpha y_\beta (\mathit{1}+y_\beta)^{-1}(\mathit{1}+c)^{-1} \xrightarrow[\tau]{}
(\mathit{1}+a)^{-1}c(\mathit{1}+c)^{-1} (\mathit{1}+b)^{-1}.
\]

On the other hand,
\begin{align*}
&(\mathit{1}+x_\alpha)^{-1} x_\alpha y_\beta (\mathit{1}+y_\beta)^{-1} (\mathit{1}+c)^{-1}
- ((\mathit{1}+a)^{-1}a)(b(\mathit{1}+b)^{-1}) (\mathit{1}+c)^{-1} \\
&= ((\mathit{1}+x_\alpha)^{-1}x_\alpha - (\mathit{1}+a)^{-1}a) y_\beta (\mathit{1}+y_\beta)^{-1}(\mathit{1}+c)^{-1} \\
&\quad + (\mathit{1}+a)^{-1} a (y_\beta (\mathit{1}+y_\beta)^{-1} -b(\mathit{1}+b)^{-1})(\mathit{1}+c)^{-1},
\end{align*}
from which, as before, we take that
\[
(\mathit{1}+x_\alpha)^{-1} x_\alpha y_\beta (\mathit{1}+y_\beta)^{-1}(\mathit{1}+c)^{-1} \xrightarrow[\tau]{}
((\mathit{1}+a)^{-1}a)(b(\mathit{1}+b)^{-1}) (\mathit{1}+c)^{-1}.
\]
Hence, we finally obtain
\begin{equation}
(\mathit{1}+a)^{-1} c(\mathit{1}+b)^{-1} = ((\mathit{1}+a)^{-1}a)(b(\mathit{1}+b)^{-1}).
\end{equation}

Suppose now that two other nets $\{ x'_\lambda \}, \{y'_\mu\}$ exist in $(\A_0)_+$ such that
\[
x'_\lambda \xrightarrow[\tau]{} a, \  y'_\mu \xrightarrow[\tau]{} b \text{ and } x'_\lambda y'_\mu
\xrightarrow[\tau]{} c'.
\]
Working exactly as before we come to the equality
\[
(\mathit{1}+a)^{-1} c'(\mathit{1}+b)^{-1} = ((\mathit{1}+a)^{-1}a) (b(\mathit{1}+b)^{-1}),
\]
which together with (6.1) gives
\[
(\mathit{1}+a)^{-1} c(\mathit{1}+b)^{-1} = (\mathit{1}+a)^{-1} c'(\mathit{1}+b)^{-1} \Leftrightarrow c=c'.
\]
\end{proof}

We may now set the following

\begin{definition}{\rm
Let $a,b \in \A[\tau]_{q+}$ with $a \in L(b)$ and $x,y \in \A_0$.
The product of the elements $ax, by$ is defined as follows:
\[
(ax)(by):= (ab)xy.
\]
}\end{definition}

Further, we consider the spectrum of an element $a \in
\A[\tau]_{q+}$.

\begin{definition}{\rm
Let $a \in \A[\tau]_{q+}$. The \textit{spectrum} of $a$ denoted by
$\sigma_{\A_0}(a)$, is that subset of $\mathbb{C}^*$, defined in the
following way:

$\bullet$ \ \ Let $\lambda \in \mathbb{C}$. Then $ \lambda \in
\sigma_{\A_0}(a) \Leftrightarrow \lambda \mathit{1} -a$ has no
inverse in $\A_0$;

$\bullet$ \ \ $\infty \in \sigma_{\A_0}(a) \Leftrightarrow a \not \in \A_0$.
}\end{definition}

\begin{lemma}
Let $a \in \A[\tau]_{q+}$. Then,
\[
\sigma_{\A_0}(a)= \{ \hat{a}(\varphi): \varphi \in \M(\A_0) \} \subset \mathbb{R}_+ \cup \{ \infty \}.
\]
In particular, $\sigma_{\A_0}(a)$ is a locally compact subset of $\mathbb{C}^*$.
\end{lemma}
\begin{proof}
Let $\lambda \in \mathbb{C}$. Then (also see Theorem 5.8),
\[
\lambda \not \in \sigma_{\A_0}(a) \Leftrightarrow (\lambda
\mathit{1} -a)^{-1} \in \A_0 \Leftrightarrow \lambda \neq
\hat{a}(\varphi), \forall \ \varphi \in \M(\A_0).
\]
Let now $\lambda = \infty$. Then,
\begin{align*}
\lambda \in \sigma_{\A_0}(a) \Leftrightarrow a \not \in \A_0
& \Leftrightarrow \hat{a} \not \in \mathcal{C}(\M(\A_0)) \\
& \Leftrightarrow \exists \ \varphi_0 \in \M(\A_0): \hat{a} (\varphi_0)=\infty.
\end{align*}
The rest is clear.
\end{proof}

If $a \in \A[\tau]_{q+}$, denote by $\mathcal{C}_b
(\sigma_{\A_0}(a))$, the $C^*$-algebra of all bounded continuous
functions on $\sigma_{\A_0}(a)$. For $n \in \mathbb{N}$ and $f \in
\mathcal{C}(\sigma_{\A_0}(a))$, define the function
\begin{equation}
g_n (\lambda): = \frac{f(\lambda)}{(\mathit{1}+\lambda)^n}, \quad \lambda \in \sigma_{\A_0}(a).
\end{equation}
In this regard, set
\begin{equation}
C_n(\sigma_{\A_0}(a)):= \{ f \in C(\sigma_{\A_0}(a) \cap \mathbb{R}): g_n \in C_b(\sigma_{\A_0}(a)) \}.
\end{equation}
Then,
\[
C_b(\sigma_{\A_0}(a)) \subset C_1(\sigma_{\A_0}(a)) \subset C_2 (\sigma_{\A_0}(a)) \subset \cdots .
\]
Now, the promised functional calculus for quasi-positive elements in
$\A[\tau]$ is given by the following

\begin{theorem}
Let $a \in \A[\tau]_{q+}$. Suppose that the element $a^n$ is
well-defined for some $n \in \mathbb{N}$. Then, there is a unique
$*$-isomorphism $f \mapsto f(a)$ from $\bigcup\limits^n_{k=1} C_k
(\sigma_{\A_0}(a))$ into $\A[\tau]$, in such a way that:

\emph{(i)} If $u_0 \in \bigcup\limits^n_{k=1} C_k (\sigma_{\A_0}(a))$, with $u_0(\lambda)= \mathit{1}$,
for each $\lambda \in \sigma_{\A_0}(a)$, then $u_0(a) =  \mathit{1} \in \A_0 \hookrightarrow \A[\tau]$.

\emph{(ii)} If $u_1 \in \bigcup\limits^n_{k=1} C_k (\sigma_{\A_0}(a))$, with $u_1(\lambda)=\lambda$,
for each $\lambda \in \sigma_{\A_0}(a)$, then $u_1(a) = a \in \A[\tau]$.

\emph{(iii)} $\widehat{f(a)}(\varphi)= f(\hat{a}(\varphi))$, for any $f \in \bigcup\limits^n_{k=1} C_k
(\sigma_{\A_0}(a))$ and $\varphi \in \M(\A_0)$;

\emph{(iv)} $(f_1+f_2)(a)= f_1(a)+f_2(a)$, for any $f_1, f_2 \in \bigcup\limits^n_{k=1}
C_k (\sigma_{\A_0}(a))$, \\
\hspace*{12mm} $(\lambda f)(a) = \lambda f(a)$, for any $f \in \bigcup\limits^n_{k=1}
C_k (\sigma_{\A_0}(a))$ and $\lambda \in \mathbb{C}$, \\
\hspace*{12mm} $(f_1 f_2)(a)= f_1(a) f_2(a)$, for any $f_j \in
C_{k_j}(\sigma_{\A_0}(a)), \ j=1,2$, with $k_1+k_2 \leq n$.

\emph{(v)} Restricted to $C_b(\sigma_{\A_0}(a))$ the map $f \mapsto
f(a)$ is an isometric $*$-isomorphism of the $C^*$-algebra
$C_b(\sigma_{\A_0}(a))$ onto the closed $*$-subalgebra of the
$C^*$-algebra $\A_0$ generated by $\mathit{1}$ and
$(\mathit{1}+a)^{-1}$.
\end{theorem}
\begin{proof}
Let $f \in \bigcup\limits^n_{k=1} C_k (\sigma_{\A_0}(a))$. Then, $f
\in C_k (\sigma_{\A_0}(a))$, for some $k$ with $1 \leq k \leq n$,
and $g_k \in C_b (\sigma_{\A_0}(a))$ with $g_k(\lambda ):=
\frac{f(\lambda)}{(\mathit{1}+\lambda)^k}, \lambda \in
\sigma_{\A_0}(a)$. From Lemma 6.5 we have that $g_k \circ \hat{a}
\in C(\M(\A_0))$, therefore (Gel'fand-Naimark theorem) there is a
unique element $g_k(a) \in \A_0$ such that
\begin{equation}
\widehat{g_k(a)} (\varphi)= g_k(\hat{a}(\varphi)), \quad \forall \
\varphi \in \M(\A_0).
\end{equation}
Now let
\begin{equation}
f(a):= g_k(a) (\mathit{1}+a)^k \in \A[\tau].
\end{equation}
We shall show that $f(a)$ does not depend on $k$, $1 \leq k \leq n$. Indeed, let $f \in C_j(\sigma_{\A_0}(a))$ with:

$\bullet$ \ \ $j \leq k$; then for each $\lambda \in \sigma_{\A_0}(a)$,
\[
g_k(\lambda)= \frac{f(\lambda)}{(\mathit{1}+\lambda)^k}= \frac{f(\lambda)}{(\mathit{1}+\lambda)^j}
\frac{1}{(\mathit{1}+\lambda)^{k-j}}= g_j(\lambda) \frac{1}{(\mathit{1}+\lambda)^{k-j}}.
\]
Hence, $g_k(a)= g_j(a)(\mathit{1}+a)^{-(k-j)} \in \A_0 \text{ and } $
\begin{equation}
g_k(a) (\mathit{1}+a)^k = g_j(a) (\mathit{1}+a)^j;
\end{equation}

$\bullet$ \ \ $j>k$; in this case too, one takes (6.6) in a similar way. So, the element
$f(a) \in \A[\tau]$ is well-defined by (6.5). Now, it is easily seen that the map
\[
f \mapsto f(a) \text{ from } \bigcup\limits^n_{k=1} C_k (\sigma_{\A_0}(a)) \text{ into } \A[\tau]
\]
is a $*$-isomorphism with the properties (i), (ii), (iii).

(iv) Consider the functions $f_1 \in C_{k_1}(\sigma_{\A_0}(a))$,
$f_2 \in C_{k_2}(\sigma_{\A_0}(a))$ with $k_1+k_2 \leq n$. Then (see
(6.3) and discussion before (6.4)), $g_{k_i} \in
C_b(\sigma_{\A_0}(a))$ with $g_{k_i}(a)$ unique in $\A_0$, $i=1,2$.
Define the function $f(\lambda):= f_1(\lambda) f_2(\lambda)$,
$\lambda \in \sigma_{\A_0}(a)$. Then, $f \in
C_{k_1+k_2}(\sigma_{\A_0}(a))$ and
\[
g_{k_1+k_2}(\lambda) = \frac{f(\lambda)}{(\mathit{1}+\lambda)^{k_1+k_2}}=  g_{k_1}(\lambda) g_{k_2}(\lambda),
\quad \lambda \in \sigma_{\A_0}(a),
\]
that is $g_{k_1+k_2} \in C_b(\sigma_{\A_0}(a))$. Thus, $g_{k_1+k_2}(a) = g_{k_1}(a) g_{k_2}(a) \in \A_0$.
Moreover (see also Definition 6.3 and (6.5))
\begin{align*}
(f_1f_2)(a)=f(a)
&= g_{k_1+k_2}(a) (\mathit{1}+a)^{k_1+k_2} \\
&= (g_{k_1}(a)(\mathit{1}+a)^{k_1}) (g_{k_2}(a)(\mathit{1}+a)^{k_2}) \\
&= f_1(a) f_2(a).
\end{align*}
The first two equalities in (iv) are similarly shown.

(v) Arguing as in (6.4) and taking into account Lemma 6.5, we easily reach at the conclusion
\end{proof}

\begin{corollary}
Let $a \in \A[\tau]_{q+}$ and $n \in \mathbb{N}$. Then, there is a
unique $b \in \A[\tau]_{q+}$ such that $a=b^n$. The element $b$ is
called quasi $n$th-root of a and is denoted by $a^{\frac{1}{n}}$.
If, in particular, $n=2$, the element $a^{\frac{1}{2}}$ is called
quasi square-root of $a$.
\end{corollary}
\begin{proof}
Consider the functions $f_1(\lambda) := \lambda^\frac{1}{n}$ and $f_2(\lambda):= \lambda^{1-\frac{1}{n}}$,
$\lambda \geq 0$, which clearly belong to $C_1(\sigma_{\A_0}(a))$. Then (see (6.2), (6.3)),
$g_1,g_2 \in C_b(\sigma_{\A_0}(a))$ with $g_1(\lambda)= f_1(\lambda)(\mathit{1}+\lambda)^{-1}$,
$g_2(\lambda)=f_2(\lambda)(\mathit{1}+\lambda)^{-1}$, $\lambda \geq 0$. Theorem 6.6 gives that
the elements $f_1(a), f_2(a)$ are uniquely defined in $\A[\tau]$ with
\[
f_1(a)= g_1(a) (\mathit{1}+a), \quad f_2(a) =g_2(a)(\mathit{1}+a),
\]
where $g_i(a) \in (\A_0)_+$, $i=1,2$ (see, e.g., (6.4)). Moreover
(also see Proposition 3.2, (1) and (2)), for each $\varepsilon >0$
\begin{align*}
&(\A_0)_+ \ni g_1(a) (\mathit{1}+a)(\mathit{1}+\varepsilon a)^{-1} \xrightarrow[\tau]{\varepsilon
\rightarrow 0} f_1(a), \text{ resp. } \\
&(\A_0)_+ \ni g_2(a) (\mathit{1}+a)(\mathit{1}+\varepsilon a)^{-1}
\xrightarrow[\tau]{\varepsilon \rightarrow 0} f_2(a).
\end{align*}

On the other hand, since $(f_1 f_2) (\lambda)= \lambda$, from
Theorem 6.6,(ii) we have that $(f_1 f_2)(a)=a$, therefore (also see
Proposition 3.2, (2))
\[
(g_1(a)(\mathit{1}+a)(\mathit{1}+\varepsilon a)^{-1}) (g_2(a) (\mathit{1}+a)(\mathit{1}+\varepsilon a)^{-1} )
= a(\mathit{1}+\varepsilon a)^{-1} \xrightarrow[\tau]{\varepsilon \rightarrow 0} a.
\]
So, from Definition 6.1, we conclude that
\[
f_1(a) \in L(f_2(a)) \text{ and } a= f_1(a) f_2(a).
\]
Now, since $f_2(a) \in \A[\tau]_{q+}$, we repeat the previous
procedure with $f_2(a)$ in the place of $a$, so that continuing in
this way we finally obtain
\[
a=f_1(a)f_1(a) \cdots f_1(a) \quad (n \text{-times}).
\]
The proof is completed by taking $b=f_1(a)$.
\end{proof}

\section{Structure of noncommutative locally convex quasi $C^*$-algebras}

In this Section we consider a noncommutative locally convex quasi
$C^*$-algebra $\A[\tau]$ over a unital $C^*$-algebra $\A_0$ and we
investigate the following: (a) Conditions under which such an
algebra is continuously embedded in a locally convex quasi
$C^*$-algebra of operators (Theorems 7.3, 7.5); (b) a functional
calculus for the commutatively quasi-positive elements in $\A[\tau]$
(Theorem 7.8).

\begin{definition}{\rm
Let $\D$ be a dense subspace of a Hilbert space $\H$. A \textit{$*$-representation} $\pi$ of
$\A[\tau]$ is a linear map from $\A$ into $\L^\dag(\D, \H)$ (see beginning of Section 4)
with the following properties:

(i) $\pi$ is a $*$-representation of $\A_0$;

(ii) $\pi(a)^\dag = \pi(a^*), \forall \ a \in \A$;

(iii) $ \pi(ax)= \pi(a) \Box \pi(x)$ and $\pi(xa)= \pi(x) \Box
\pi(a), \forall \ a \in \A$ and $x \in \A_0$, where $\Box$ is the
(weak) partial multiplication in $\L^\dag(\D, \H)$ (ibid.) Having a
$*$-representation $\pi$ as before, we write $\D(\pi)$ in the place
of $\D$ and $\H_\pi$ in the place of $\H$. By a \textit{$(\tau,
\tau_{s^*})$-continuous $*$-representation} $\pi$ of $\A[\tau]$, we
clearly mean continuity of $\pi$, when $\L^\dag(\D(\pi), \H_\pi)$
carries the locally convex topology $\tau_{s^*}$ (see Section 4). }
\end{definition}

\begin{lemma}
Let $\pi$ be a $*$-representation of $\A[\tau]$ with domain $\D(\pi)$ dense in $\H_\pi$.
Let also $\B$ be an admissible subset of $\B(\pi(\A))$.
The following hold:

\emph{(1)} If $\pi$ is $(\tau, \tau_{s^*})$-continuous, then $\pi(\A)[\tau_{s^*}]$ is
a locally convex quasi $C^*$-algebra over the $C^*$-algebra $\pi(\A_0)$.

\emph{(2)} If $\pi$ is $(\tau, \tau^u_*(\B))$-continuous \emph{(in
the spirit of Definition 7.1)}, then $\pi(\A)[\tau^u_*(\B)]$ is a
locally convex quasi $C^*$-algebra over $\pi(\A_0)$.
\end{lemma}
\begin{proof}
Clearly $\pi(\A_0)$ is a $C^*$-algebra and
\[
\pi: \A[\tau] \rightarrow \pi(\A) [\tau_{s^*}] \subset \widetilde{\pi(\A_0)}[\tau_{s^*}]
\]
is a $(\tau, \tau_{s^*})$-continuous $*$-representation of $\A[\tau]$, with $\pi(\A)$ a
quasi $*$-algebra over $\pi(\A_0)$ and $\widetilde{\pi(\A_0)}[\tau_{s^*}]$ (similarly
$\widetilde{\pi(\A_0)}[\tau^u_*(\B)]$) a locally convex quasi $C^*$-algebra over $\pi(\A_0)$.
So, (1) and (2) follow from Definition 3.3.
\end{proof}

Now, a sesquilinear form $\varphi$ on $\A \times \A$ is called
\textit{positive}, resp. \textit{invariant}, if and only if
$\varphi(a,a) \geq 0$, for each $a \in \A$, resp. $\varphi(ax,y)=
\varphi(x,a^*y)$, for all $a \in \A$ and $x,y \in \A_0$. Moreover,
$\varphi$ is called \textit{$\tau$-continuous}, if $| \varphi(a,b)|
\leq p(a) p(b)$ for some $\tau$-continuous seminorm $p$ on $\A$ and
all $a,b \in \A$.

Further, let $\varphi$ be a $\tau$-continuous positive invariant
sesquilinear form on $\A_0 \times \A_0$. Then, $\tilde{\varphi}$
denotes the extension of $\varphi$ to a $\tau$-continuous positive
invariant sesquilinear form on $\A \times \A$. Moreover, let
$(\pi_\varphi, \lambda_\varphi, \H_\varphi)$ be the
$GNS$-construction for $\varphi$ (see, for instance, \cite[Section
9.1]{ait3}). Then, $\pi_\varphi$ is extended on $\A$, as follows:
\begin{equation}
\pi_\varphi(a) \lambda_\varphi(x):=\displaystyle \lim_\alpha
\pi_\varphi(x_\alpha) \lambda_\varphi(x), \quad \forall \ x \in
\A_0,
\end{equation}
where $\{x_\alpha\}$ is a net in $\A[\tau]$ with $a= \tau$-$\displaystyle \lim_\alpha x_\alpha$.
By the very definitions and the $\tau$-continuity of $\varphi$, it follows that $\pi_\varphi$ is
a $(\tau, \tau_{s^*})$-continuous $*$-representation of $\A$. Now, put
\[
\mathcal{S}(\A_0):= \{ \tau \text{-continuous positive invariant sesquilinear forms } \varphi
\text{ on } \A_0 \times \A_0 \}.
\]
We shall say that the set $\mathcal{S}(\A_0)$ is
\textit{sufficient}, whenever
\[
a \in \A \text{ with } \tilde{\varphi}(a,a)=0, \forall \ \varphi \in
\mathcal{S}(\A_0), \text{ implies } a=0.
\]
From the results that follow, Theorems 7.3, 7.5 (and, of course,
Corollary 7.4) give answers to the question (a) stated at the
beginning of this Section. These results can be viewed as analogues
of the Gel'fand-Naimark theorem, in the case of locally convex quasi
$C^*$-algebras.
\begin{theorem}
Let $\A[\tau]$ be a locally convex quasi $C^*$-algebra over a unital $C^*$-algebra $\A_0$.
The following statements are equivalent:

\emph{(1)} There exists a faithful $(\tau, \tau_{s^*})$-continuous $*$-representation $\pi$ of $\A$.

\emph{(2)} The set $\mathcal{S}(\A_0)$ is sufficient.
\end{theorem}
\begin{proof}
(1) $\Rightarrow$ (2) For every $\xi \in \D(\pi)$ define
\[
\varphi_\xi (x,y) := (\pi(x) \xi | \pi(y) \xi), \quad \forall \ x, y
\in \A_0.
\]
Then, $\{ \varphi_\xi : \xi \in \D(\pi) \} \subset
\mathcal{S}(\A_0)$, so that from the preceding discussion it follows
easily that $\mathcal{S}(\A_0)$ is sufficient.

(2) $\Rightarrow$ (1) Let $\varphi \in \mathcal{S}(\A_0)$ and
$(\pi_\varphi, \lambda_\varphi, \H_\varphi)$ the $GNS$-construction
for $\varphi$. Then, as we noticed before (see (7.1)), $\pi_\varphi$
extends to a $(\tau, \tau_{s^*})$-continuous $*$-representation of
$\A$ with $\D(\pi_\varphi) = \lambda_\varphi(\A_0)$. Now, take
\begin{align*}
\D(\pi):= & \left\{ (\lambda_\varphi(x_\varphi))_{\varphi \in \mathcal{S}(\A_0)} \right.
\in \bigoplus\limits_{\varphi \in \mathcal{S}(\A_0)} \H_\varphi: x_\varphi \in \A_0 \text{ and } \\
& \left. \quad \lambda_\varphi(x_\varphi)=0, \text{ except for a finite number of }
\varphi\text{'s from } \mathcal{S}(\A_0) \right\}
\end{align*}
and define
\[
\pi(a) (\lambda_\varphi(x_\varphi)):= (\lambda_\varphi(ax_\varphi)),
\forall \ a \in \A \text{ and } (\lambda_\varphi(x_\varphi)) \in
\D(\pi).
\]
Then, it is easily seen that $\pi$ is a faithful $(\tau, \tau_{s^*})$-continuous $*$-representation of $\A$.
\end{proof}

Results for quasi $*$-algebras over a unital $C^*$-algebra $\A_0$
related to Theorem 7.3, have been considered in \cite[Theorem
3.3]{ba-maria-i-t} and \cite[Theorem 3.2]{maria-i-kue}.

Now an application of Theorem 7.3 and Lemma 7.2, gives the following
\begin{corollary}
Let $\A[\tau]$, $\A_0$ be as in Theorem \emph{7.3}. Suppose that the
set $\mathcal{S}(\A_0)$ is sufficient. Then, the locally convex
quasi $C^*$-algebra $\A[\tau]$ over $\A_0$ is continuously embedded
in a locally convex quasi $C^*$-algebra of operators.
\end{corollary}

The next theorem gives further conditions under which a locally convex quasi
$C^*$-algebra $\A[\tau]$ can be continuously embedded in a locally convex quasi $C^*$-algebra of operators.

\begin{theorem}
Let $\A[\tau]$ be a locally convex quasi $C^*$-algebra over $\A_0$. Suppose
the multiplication of $\A_0$ satisfies
the following condition:

For every $\tau$-bounded subset $B$ of $\A_0$ and every $\lambda \in \Lambda$, there exist
$\lambda' \in \Lambda$ and a positive constant $c_B$ such that
\[
\sup_{y \in B} p_\lambda(xy) \leq c_B p_{\lambda'}(x), \quad \forall
\ x \in \A_0.
\]

Then, the next statements are equivalent:

\emph{(i)} There is a faithful $(\tau, \tau^u_*(\B))$-continuous $*$-representation
$\pi$ of $\A$, where $\B$ is an admissible subset of $\B(\pi(\A))$.

\emph{(ii)} There is a faithful $(\tau, \tau_{s^*})$-continuous $*$-representation of $\A$.

\emph{(iii)} The set $\mathcal{S}(\A_0)$ is sufficient.
\end{theorem}
\begin{proof}
(i) $\Rightarrow$ (ii) It is trivial (see (4.3)).

(ii) $\Rightarrow$ (iii) It follows from Theorem 7.3.

(iii) $\Rightarrow$ (i) Let $\varphi \in \mathcal{S}(\A_0)$ and
$(\pi_\varphi, \lambda_\varphi, \H_\varphi)$ the $GNS$-construction
for $\varphi$ (see discussion before Theorem 7.3). Set
\[
\B_\varphi := \{ \lambda_\varphi(B): B \text{ a } \tau \text{-bounded subset of } \A_0 \}.
\]
Then, for each $\tau$-bounded subset $B$ of $\A_0$, we have
\[
\sup_{y \in B} \| \pi_\varphi(a)  \lambda_\varphi(y) \| = \sup_{y \in B} \varphi(ay, ay)^{1/2}
\leq \sup_{y \in B} p_\lambda (ay) \leq c_B p_{\lambda'}(a),
\]
for all $a \in \A$ and some $\lambda, \lambda' \in \Lambda$. It is clear now that
$\lambda_\varphi(B) \in \B(\pi_\varphi(\A))$ and that (see (4.2)) $\pi_\varphi$ is
$(\tau, \tau^u_*(\B_\varphi))$-continuous.
Let now $\pi$ be as in the proof of Theorem 7.3. Put
\[
\B_\pi:= \{ \bigoplus\limits^{\text{finite}}_{\varphi \in \mathcal{S}(\A_0)}
\lambda_\varphi(B_\varphi): B_\varphi \text{ a } \tau\text{-bounded subset of } \A_0 \}.
\]
Then, it is easily seen that $\B_\pi$ is an admissible subset of
$\B(\pi(\A))$ and $\pi$ a faithful $(\tau,
\tau^u_*(\B_\pi))$-continuous $*$-representation of $\A$.
\end{proof}

An analogue of Corollary 7.4 is stated in the case of Theorem 7.5, too.

\medskip

Taking again $\A[\tau], \A_0$ as in Theorem 7.3, we proceed to the
study of a functional calculus for the commutatively quasi-positive
elements of $\A[\tau]$ (see (b) at the beginning of this Section).
So, let $a \in \A[\tau]_{cq+}$. Then, from Proposition 3.2,(1), the
element $(\mathit{1}+a)^{-1}$ exists and belongs to
$\mathcal{U}(\A_0)_+$. Consider the maximal commutative
$C^*$-subalgebra $C^*(a)$ of $\A_0$ containing the elements
$\mathit{1}, (\mathit{1}+a)^{-1}$. Then,
\begin{enumerate}
\item[$\bullet$] $C^*(a)[\tau]$ \textit{satisfies the properties}
(T$_1$)-(T$_4$) of Section 3. The properties (T$_1$)-(T$_3$) are
trivially checked. We must check (T$_4$).
\end{enumerate}
First we prove that $\mathcal{U}(C^*(a))_+$ is $\tau$-closed. Let
$\{x_\alpha \}$ be a net in $\mathcal{U}(C^*(a))_+$ such that
$x_\alpha \xrightarrow[\tau]{} x$. But, $\mathcal{U}(C^*(a))+
\subset \mathcal{U}(\A_0)_+$ and since $\mathcal{U}(\A_0)_+$ is
$\tau$-closed we have that $x \in \mathcal{U}(\A_0)_+$. On the other
hand,
\[
xy \xleftarrow[\tau]{} x_\alpha y = y x_\alpha \xrightarrow[\tau]{}
yx, \quad \forall \ y \in C^*(a).
\]
Hence, $xy = yx$, which by the maximality of $C^*(a)$ means that $x
\in C^*(a)$ and finally $x \in \mathcal{U}(C^*(a))_+$. Thus,
$\mathcal{U}(C^*(a))+$ is $\tau$-closed. Now, take an arbitrary $x
\in \widetilde{C^*(a)}[\tau]_{q+} \cap C^*(a)$. Then, $x \in
\A[\tau]_{q+} \cap \A_0 =(\A_0)_+ $, and so $x \in C^*(a) \cap
(\A_0)_+ = C^*(a)_+ $. This completes the proof of (T$_4$). Thus,
the following is proved:

\begin{proposition}
Let $\A[\tau]$ be a locally convex quasi $C^*$-algebra over a unital
$C^*$-algebra $\A_0$. Let $a \in \A[\tau]_{cq+}$ and $C^*(a)$ the
maximal commutative $C^*$-subalgebra of $\A_0$ containing $\{
\mathit{1}, (\mathit{1}+a)^{-1}\}$. Then, $\widetilde{C^*(a)}[\tau]$
is a commutative locally convex quasi $C^*$-algebra over $C^*(a)$.
\end{proposition}

\begin{corollary}
The element $a$ belongs to $\widetilde{C^*(a)}[\tau]_{q+}$.
\end{corollary}
\begin{proof}
Since $a \in \A[\tau]_{cq+}$, Proposition 3.2,(2) implies that for
every $\varepsilon>0$, $a (\mathit{1}+\varepsilon
a)^{-1}=\frac{1}{\varepsilon} \left(
\mathit{1}-(\mathit{1}+\varepsilon a)^{-1} \right) \in (\A_0)_+$.
Now, since $(\mathit{1}+a)^{-1}$ commutes with every element $\omega
\in C^*(a)$, it follows that $\omega$ also commutes with
$\mathit{1}+a$, hence with $a$, therefore with
$(\mathit{1}+\varepsilon a)^{-1}$ too. Thus,
$a(\mathit{1}+\varepsilon a)^{-1} \in C^*(a)$, for each $\varepsilon
> 0$. Since moreover, $a=\tau$-$\displaystyle \lim_{\varepsilon
\rightarrow 0}  a(\mathit{1}+ \varepsilon a)^{-1}$ (ibid.),
Definition 3.1 gives that $a \in \widetilde{C^*(a)}[\tau]_{q+}$.
\end{proof}

It is now clear from Corollary 7.7 that making use of Theorem 6.6
for $\widetilde{C^*(a)}[\tau]_{q+}$, we can obtain the promised
functional calculus for the commutatively quasi-positive elements of
the noncommutative locally convex quasi $C^*$-algebra $\A[\tau]$.
That is, we have the following

\begin{theorem}
Let $\A[\tau]$ be a noncommutative locally convex quasi
$C^*$-algebra over a unital $C^*$-algebra $\A_0$. Let $a \in
\A[\tau]_{cq+}$ such that $a^n$ is well defined for some $n \in
\mathbb{N}$. Then, there is a unique $*$-isomorphism $f \mapsto
f(a)$ from $\bigcup\limits^n_{k=1} C_k(\sigma_{C^*(a)}(a))$ into
$\A[\tau]$ such that:

\emph{(1)} If $u_0 \in \bigcup\limits^n_{k=1}
C_k(\sigma_{C^*(a)}(a))$ with $u_0(\lambda)=1$, for each $\lambda
\in \sigma_{C^*(a)}(a)$, then $u_0(a) =\mathit{1} \in C^*(a)
\hookrightarrow \A[\tau]$.

\emph{(2)} If $u_1 \in \bigcup\limits^n_{k=1}
C_k(\sigma_{C^*(a)}(a))$ with $u_1(\lambda)= \lambda$, for each
$\lambda \in \sigma_{C^*(a)}(a)$, then $u_1(a)= a \in \A[\tau]$.

\emph{(3)} $\widehat{f(a)}(\varphi)= f(\hat{a}(\varphi))$, for any $f \in \bigcup\limits^n_{k=1}
C_k(\sigma_{C^*(a)}(a)) $ and $ \varphi \in \M(C^*(a))$.

\emph{(4)} $(f_1+f_2)(a) =f_1(a)+f_2(a)$, for any $f_1,  f_2 \in \bigcup\limits^n_{k=1}C_k(\sigma_{C^*(a)}(a))$, \\
\hspace*{12mm} $(\lambda f)(a)= \lambda f(a)$, for any $f \in \bigcup\limits^n_{k=1}C_k(\sigma_{C^*(a)}(a))$
and $\lambda \in \mathbb{C}$,
\\
\hspace*{12mm} $(f_1f_2)(a)= f_1(a)f_2(a)$, for any $f_j \in
C_{k_j}(\sigma_{C^*(a)}(a)), \ j=1,2$, with $k_1+k_2 \leq n$.

\emph{(5)} Restricted to $C_b(\sigma_{C^*(a)}(a))$ the map $f
\mapsto f(a)$ is an isometric $*$-isomorphism of the $C^*$-algebra
$C_b(\sigma_{C^*(a)}(a))$ onto the closed $*$-subalgebra of the
$C^*$-algebra $C^*(a)$ generated by $\mathit{1}$ and
$(\mathit{1}+a)^{-1}$.
\end{theorem}

Now, an application of Corollary 6.7 for the commutative locally
convex quasi $C^*$-algebra $\widetilde{C^*(a)}[\tau]$ and Theorem
7.8 give the following

\begin{corollary}
Let $\A[\tau]$, $\A_0$ be as in Theorem \emph{7.8}. Let $a \in
\A[\tau]_{cq+}$ and $n \in \mathbb{N}$. Then, there is a unique
element $b \in \A[\tau]_{cq+}$ such that $a=b^n$. The element $b$ is
called commutatively quasi $n$th-root of $a$ and is denoted by
$a^{\frac{1}{n}}$. If $n=2$, the element $a^{\frac{1}{2}}$ is called
commutatively quasi square root of $a$.
\end{corollary}

{\bf Acknowledgment.} This work begun during the stay of two of us
(F.B and C.T.) at the Department of Applied Mathematics, Fukuoka
University, in the summer of 2005 and it was concluded during the
3-month stay of the second author (M.F.) at the same Department in
the spring of 2006. All three of us gratefully acknowledge the
financial support of this Department and the warm hospitality of its
staff.

\begin{flushleft}
Dipartimento di Matematica ed Applicazioni, Fac Ingegneria,
Universita di\\ Palermo, I-90128 Palermo, Italy\\
E-mail address: bagarell@unipa.it
\end{flushleft}

\begin{flushleft}
Department of Mathematics, University of Athens, Panepistimiopolis,\\ Athens
15784, Greece\\ E-mail address: fragoulop@math.uoa.gr
\end{flushleft}

\begin{flushleft}
Department of Applied Mathematics, Fukuoka University,\\ Fukuoka 814-0180, Japan\\
E-mail address: a-inoue@fukuoka-u.ac.jp
\end{flushleft}

\begin{flushleft}
Dipartimento di Matematica ed Applicazioni, Universita
di Palermo,\\ I-90123 Palermo, Italy\\
E-mail address: trapani@unipa.it
\end{flushleft}

\end{document}